%% file: loveQFT.tex
\documentclass[11pt]{article} 

\usepackage[utf8]{inputenc} 

\usepackage{geometry} 
\usepackage{graphicx} 

\usepackage{booktabs} 
\usepackage{array} 
\usepackage{paralist} 
\usepackage{verbatim} 
\usepackage{subfig} 
\usepackage{amsmath}
\usepackage{amssymb}
\usepackage{amsfonts}
\usepackage{bbm}
\usepackage[sectionbib]{chapterbib}
\usepackage{nicefrac}
\usepackage{dsfont}

\newtheorem{theorem}{Theorem}

\newtheorem{definition}[theorem]{Definition}

\usepackage{hyperref}

\usepackage{fancyhdr} 
\usepackage{sectsty}
\allsectionsfont{\sffamily\mdseries\upshape} 

\usepackage[nottoc,notlof,notlot]{tocbibind} 
\usepackage[titles,subfigure]{tocloft} 

\def\MC{{\mathbbm C}}
\def\MR{{\mathbbm R}}

\def\tr{\operatorname{tr}}
\def\MZ{{\mathbbm Z}}
\def\MN{{\mathbbm N}}

\providecommand{\abs}[1]{\ensuremath{\left|#1\right|}}

\begin{document}
\title{How I Learned to Stop Worrying and Love QFT\\LMU, Summer 2011}
\author{Robert C. Helling\\{\small (helling@atdotde.de)}\\ Notes by Constantin Sluka and Mario Flory}
\date{August 4, 2011} 
\maketitle
\abstract{Lecture notes of a block course explaining why quantum field
  theory might be in a better mathematical state than one gets the
  impression from the typical introduction to the topic. It is
  explained how to make sense of a perturbative expansion that fails
  to converge and how to express Feynman loop integrals and their
  renormalization using the language of distribtions rather than
  divergent, ill-defined integrals.}

\setcounter{tocdepth}{1}
{\def\large{\small}
}
\input introduction
\vfill\eject
\input perturbation
\input distributions

\input conclusions
\section*{Acknowledgements} 
A lot of the material presented we learned from Klaus Fredenhagen,
Dirk Prange and Marcel Vonk. We would like to thank the
Elitemasterprogramme ``Theoretical and Mathematical Physics'' and
Elitenetzwerk Bayern.

\bibliographystyle{JHEP}
\bibliography{loveQFT}
\end{document}

%% file: introduction.tex
\section{Introduction}

Physicists are often lax when it comes to mathematical rigor and use
objects that do not exist according to strict mathematical standards
or happily exchange limits without justification. This different
culture of ``everything is allowed as long as it is not proven to be
wrong and even then it sometimes ok because we do not actually mean
what we are writing'' is preferred by many as it allows to ``focus on
the content rather than the formal aspects'' and to progress at a much
faster pace.

This attitude can be seen when physicists talk about quantum mechanics
and treat operators as if they were matrices and plane waves as if
they are elements of the relevant Hilbert space. This is generally
accepted since one has the feeling that these arguments can easily be
repaired at the expense of clarity by talking about wave packets
instead of plane waves and (like it is discussed at length in our
``Mathematical Quantum Mechanics'' course) by talking about quadratic
forms instead of the operators directly.

The situation appears to be very different in the case of quantum
field theory: There, most of the time, one deals with perturbative
series expansions in the coupling constant without thinking about
convergence (or if one spends some thought on this one easily sees
that the radius of convergence has to be zero) and the individual
terms in the series turn out to be divergent and one obtains
reasonable, finite expressions after some very doubtful formal
manipulations (often presented as subtracting infinity from infinity
in the ``right way''). The typical QFT course, unlike quantum
mechanics above, does not indicate any way to ``repair'' these
mathematical shortcomings. Often, one is left with the impression that
there is some blind faith required on the side of the physicists or at
least that some black magic is helping to obtain numerical values that
fit so impressively what is measured in experiments from very doubtful
expressions.

In these notes we will indicate some ways in which these treatments can
be made more exact mathematically thus providing some cure to the
mathematical uneasiness related to quantum field theory. In
particular, we will argue that QFT is not ``obviously wrong'' as claimed
by some mistakenly confusing mathematical rigor with correctness.

Concretely, we want to explain how two (mostly independent) crucial
steps in QFT can be understood more mathematically:

In a simplified example, we will explore what conclusions can be drawn
from the perturbative expansion even though the series does not
converge for any finite value of the coupling constant. In particular
we will discuss the role of non-perturbative contributions like
instantons in the full interacting theory. We will find that up to a
certain level of accuracy (depending on the strength of coupling), the
first terms of the perturbative expansion do represent the full
answer even though summing up all terms leads to infinite, meaningless
expressions. Furthermore, at least in principle, using the technique
of ``Borel resummation'' one can express the true expression for all
values of the coupling constant in terms of just the perturbative expansion.

As a second step, at each order in perturbation theory, we will see
how by correctly using the language of distributions one can set up
the calculation of Feynman diagrams without diverging momentum
integrals. We will find that these divergences can be understood to
arise from trying to multiply distributions. We will set this up as
the problem to extend distributions from a subset of all test
functions at the expense of a finite number of undetermined quantities
that we will identify as the ``renormalized coupling
constants''. Finally we will understand how these vary when we change 
regulating functions that were introduced in the procedure which leads
to an understanding of the renormalization group in this formalism of
``causal perturbation theory''.

The aim is to argue how the techniques of physicists could be embedded
in a more mathematical language without actually doing this. At many
places we just claim results without proof or argue by analogy (for
example we will discuss a one dimensional integral instead of an
infinite dimensional path-integral). To really discuss the topic at a
mathematical level of rigor requires a lot more work and to large
extend still needs to be done for theories of relevance to particle
physics. 

All this material is not new but well known to experts in the
field. Still, we hope that these notes will be a useful complement to
standard introductions to quantum field theory for (beginning)
practitioners.


%% file: perturbation.tex
\section{Perturbative expansion --- making sense of divergent series}
Before we take a look at divergent series, we will first give a brief review of how perturbative expansion is used in quantum field theory.
\subsection{Brief overview on path integrals}
A quantum field theory in Minkowski spacetime is described by a Lagrangian density $\mathcal L(\phi,\partial\phi)$ and a generating functional of correlation functions\footnote{This subsection displays some standard expressions to set the context. For many more details see for example \cite{Osborn}.}
\begin{equation}
\label{eq:1}
\mathcal Z[J] = \int \mathcal{D}\phi e^{i\int\mathrm{d}^4x(\mathcal{L}+J\phi)}.
\end{equation}
The correlation functions can be obtained by functional derivatives of (\ref{eq:1}) with respect to $J$.
\begin{equation}
\label{eq:2}
\langle\phi(x_1)\phi(x_2)\dots\phi(x_n)\rangle = \frac{1}{\mathcal Z[0]}\left(-i\frac{\delta}{\delta J(x_1)}\right)\left(-i\frac{\delta}{\delta J(x_2)}\right)\dots\left(-i\frac{\delta}{\delta J(x_n)}\right)\mathcal Z[J] \Bigg |_{J=0}
\end{equation}
In this lecture we will use Euclidean signature for the metric instead
of Minkowski. The change between the metrics can be performed as
rotation of the time axis in the complex plane $t\rightarrow -i\tau$ if
all expressions are analytic. In Euclidean metric, the exponent in the
generating functional is real and falls of at large field values. This
gives the path integral a chance to have a mathematical definition in
terms of Wiener measures but that will not concern us in these notes.
\begin{equation}
\label{eq:3}
\mathcal Z[J] = \int \mathcal{D}\phi e^{\int\mathrm{d}^4x(\mathcal{L}+J\phi)}
\end{equation}
In general, the integral (\ref{eq:3}) cannot be computed exactly. For a scalar quantum field theory in Euclidean space the Lagrangian has the form
\begin{equation}
\mathcal{L}=\frac{1}{2}\phi(\Box-m^2)\phi - V(\phi)
\label{eq:4}
\end{equation}
with $\Box \equiv (\partial_\tau)^2+(\nabla)^2$.

If the potential $V(\phi)$ vanishes, equation (\ref{eq:3}) can be
formally computed as it becomes an integral of Gaussian type. One therefore arbitrarily splits the Lagrangian into its ``kinetic part'' $\frac{1}{2}\phi(\Box-m^2)\phi$ and its ``interaction part'' $-V(\phi)$.
\begin{align}
\mathcal Z[J] &= \int\mathcal{D}\phi e^{\frac{1}{2}\int\mathrm{d}^4x \phi(\Box-m^2)\phi}e^{-\int\mathrm{d}^4xV(\phi)}e^{-\int\mathrm{d}^4xJ\phi} \nonumber \\
&= e^{-\int\mathrm{d}^4xV(\frac{\delta}{\delta J})}\int\mathcal{D}\phi e^{\int\mathrm{d}^4x \frac{1}{2}\phi(\Box-m^2)\phi-J\phi}
\label{eq:5}
\end{align}
To obtain the Gaussian integral one has to complete the square in the exponent. This is achieved by shifting the field $\phi$:
\begin{equation}
\phi^\prime = \phi + (\Box-m^2)^{-1}J 
\label{eq:6}
\end{equation}
The inverse of $\Box-m^2$, called ``Green's function'' $G(x-y)$, is a distribution defined by
\begin{equation}
(\Box-m^2)G(x-y) = \delta(x-y).
\label{eq:7}
\end{equation}
Changing variables in the functional integral (\ref{eq:5}) leads to
\begin{equation}
\mathcal Z[J] = e^{-\int\mathrm{d}^4xV(\frac{\delta}{\delta J})}\int\mathcal{D}\phi^\prime e^{\int\mathrm{d}^4x \frac{1}{2}\phi^\prime(\Box-m^2)\phi^\prime}e^{-\int\mathrm{d}^4x\int\mathrm{d}^4y\frac{1}{2}J(x)G(x-y)J(y)}.
\label{eq:8}
\end{equation}
The complicated expression in the middle of equation (\ref{eq:8}) does
not depend on $J$ and will in fact cancel out in equation (\ref{eq:2})
for the correlation function, so we will just denote it $C$ and forget
about it:
\begin{equation*}
  \mathcal Z[J] = Ce^{-\int\mathrm{d}^4xV(\frac{\delta}{\delta J})}e^{-\int\mathrm{d}^4x\int\mathrm{d}^4y\frac{1}{2}J(x)G(x-y)J(y)}
\end{equation*}

Now let us take a look on a specific example for a quantum field theory by choosing a potential for the scalar field. We will consider our favorite $\phi^4$ theory given by the potential
\begin{equation}
V(\phi) = \lambda\phi^4.
\label{eq:9}
\end{equation}
The next step is to insert this potential in equation (\ref{eq:8}) and write the exponential as a power series in the coupling strength $\lambda$.
\begin{equation}
\label{eq:10}
\mathcal Z[J] = C\sum_{k=0}^\infty\frac{(-\lambda)^k}{k!}\int\mathrm{d}^4x_1\frac{\delta^4}{\delta J(x_1)^4}\dots\int\mathrm{d}^4x_k\frac{\delta^4}{\delta J(x_k)^4}e^{-\int\mathrm{d}^4x\int\mathrm{d}^4y\frac{1}{2}J(x)G(x-y)J(y)}
\end{equation}
We now found an expression for any general correlation function in terms of an power series expansion in the coupling strength.
\begin{align}
\langle \phi(y_1)\dots\phi(y_n)\rangle = &\frac{\delta}{\delta J(y_1)}\cdots \frac{\delta}{\delta J(y_1)} \sum_{k=0}^\infty\frac{(-\lambda)^k}{k!}\int\mathrm{d}^4x_1\frac{\delta^4}{\delta J(x_1)^4}\cdots\int\mathrm{d}^4x_k\frac{\delta^4}{\delta J(x_k)^4}  \times\nonumber \\
& e^{-\int\mathrm{d}^4x\int\mathrm{d}^4y\frac{1}{2}J(x)G(x-y)J(y)}\Bigg|_{J=0}
\label{eq:11}
\end{align}
The combinatorics of the occurring expressions in terms of integrals
over interaction points $x_i$, Green's functions and external fields can be summarized in terms of Feynman diagrams each standing for a single term in the power series in the coupling constant
$\lambda$\footnote{The careful reader wishing to avoid ill-defined expressions using path-integrals, can use this formula as the
  definition of the terms in the perturbative series.}. In the following, we want to study the convergence behavior of this power
series.

\subsection{Radius of convergence of correlation functions}
Let us briefly review the definition of the radius of convergence for a power series from introductory analyis. It is useful to think of a power series to be defined in the complex plane:
\begin{equation}
\sum_k^\infty \lambda^k(\dots) \quad \lambda\in \MC
\label{eq:12}
\end{equation} 
(If one does not like the idea of a complex coupling strength in a
quantum field theory, just restrict to the special $\lambda \in \MC$ that
happen to be real.). Every power series has a radius of convergence $R
\in [0,\infty]$ such that
\begin{equation}
\sum_k^\infty \lambda^k(\dots) 
\begin{cases}
\text{converges} & \forall \abs{\lambda} < R \\
\text{diverges} & \forall \abs{\lambda} > R.
\end{cases}
\label{eq:13}
\end{equation}

\begin{figure}[h]
  \centering
  \subfloat[$\lambda>0$]{\label{fig:potential1}\input{potential1.tex}}                
  \subfloat[$\lambda<0$]{\label{fig:potential2}\input{potential2.tex}}
  \caption{Potentials of $\phi^4$ theory}
  \label{fig:potentials}
\end{figure}
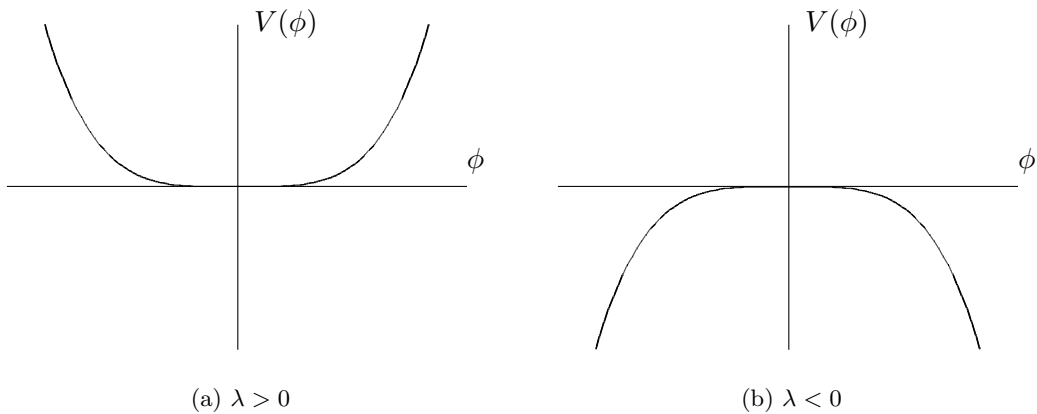

Now we want to find out the radius of convergence for the correlation
functions (\ref{eq:11}) in a quantum field theory. A physicist's
argument was given by Freeman Dyson in
1952\cite{Dyson}. Let us take a
look on the potential, for example in our $\phi^4$ theory as shown in
figure \ref{fig:potentials}. For positive coupling strength $\lambda$
the potential is bounded from below and large values of $\phi$ are
strongly disfavored. This behavior, however, gets radically different
in case of a negative $\lambda$. The potential becomes unbounded from
below and the field $\phi$ will want to run off to
$\phi=\pm\infty$. Obviously, such a behavior is highly unphysical,
since ever increasing values of $\phi$ would lead to an infinite
energy gain. It is thus clear that such a theory cannot lead to
healthy correlation functions, in other words for any negative
$\lambda$ the power series (\ref{eq:12}) will diverge\footnote{We
  expect at least a phase transition when $\lambda$ is changed from
  positive to negative values.}. From this we can conclude the radius
of convergence being $R=0$!
\begin{equation}
\sum_k^\infty\lambda^k(\dots)\qquad \text{diverges } \forall\lambda>0
\label{eq:14}
\end{equation}

For readers not satisfied by this argument using physics of unstable potentials for determining the radius of convergence, let us mention an alternative line of argument. Again, consider equation (\ref{eq:11}), this time, however, we will focus on the Feynman diagrams. At any order $k$ in the perturbation expansion there is a sum of different Feynman diagrams expressing the integrals in (\ref{eq:11}), where $k$ counts the number of vertices. The combinatorics of all Feynman diagrams shows that the number of Feynman diagrams grows like $k!$. The power series, therefore, will behave like
\begin{equation}
\sum_k^\infty \lambda^kk!(\ldots).
\label{eq:15}
\end{equation}
Assuming that $(\ldots)$ is not surprisingly suppressed for large $k$,
the coefficients of $\lambda^k$ grow faster than any power, we again
find the radius of convergence $R=0$. 

In the following, we want to give an example, how one can nevertheless
make sense of (some) divergent power series.
\subsection{Non-perturbative corrections}
In order to get a feeling for the problem of divergent power series,
we will consider a one dimensional toy problem  (rather than the
infinite dimensional problem of a path integral):
\begin{equation}
\mathcal Z(\lambda) = \int_{-\infty}^\infty\mathrm{d}x\, e^{-x^2-\lambda x^4}
\label{eq:16}
\end{equation}
We take $\lambda \geq 0$, so this integral yields some finite,
positive number. For $\lambda = 0$ the solution is well known
\begin{equation}
\mathcal Z(0) = \sqrt{\pi}.
\label{eq:17}
\end{equation}
In general, equation (\ref{eq:16}) can be expressed in terms of
special functions, e.g.\ \textit{Mathematica} gives the solution
\begin{equation}
\mathcal Z(\lambda) = \frac{e^{\frac{1}{8\lambda}}K_{1/4}({1/8\lambda})}{2\sqrt{\lambda}}
\label{eq:18}
\end{equation}
with $K_n(x)$ being the modified Bessel function of the second kind. We call solution (\ref{eq:18}) the ``full, non-perturbative answer''. Now we will do the same as in quantum field theory and split the integral into a ``kinetic'' and an ``interaction'' part, respectively. 
\subsubsection{Treating the toy model perturbatively}
Following the same procedure, we will again expand the ``interaction part'' $-\lambda x^4$ in a power series:
\begin{equation}
\mathcal Z(\lambda) = \int_{-\infty}^\infty \mathrm{d}xe^{-x^2-\lambda x^4} =  \int_{-\infty}^\infty \mathrm{d}x e^{-x^2}\sum_{k=0}^\infty \frac{(-\lambda x^4)^k}{k!}
\label{eq:19}
\end{equation}
Now comes the crucial step and ``root of all evil''. Following precisely the same steps leading towards equation (\ref{eq:11}) for correlation functions in quantum field theory, we will change the order of integration and summation, leading to the interpretation of a power series of Feynman diagrams:
\begin{equation}
\mathcal Z(\lambda) \text{``}=\text{''} \sum_{k=0}^\infty \frac{(-\lambda)^k}{k!}\int_{-\infty}^\infty \mathrm{d}x\ x^{4k}e^{-x^2}
\label{eq:20}
\end{equation}
From this step, as we will see later, the problems arise. Although
this step is forbidden (as roughly speaking, we are changing the
behavior of the integrand at $x=\pm\infty$), we are interested in to what
extent a ``perturbative solution'' obtained from equation (\ref{eq:20})
will agree with the full, non-perturbative solution
(\ref{eq:18}). Carrying on, we observe that the integral in
(\ref{eq:20}) is now of the type ``polynomial times Gaussian'' and can
be computed with standard methods. We smuggle an addtional factor $a$ into the exponent allowing us to write the integrand as derivatives of $e^{-ax^2}$
with respect to $a$ at the point $a=1$.
\begin{equation}
\mathcal Z(\lambda) = \sum_{k=0}^\infty \frac{(-\lambda)^k}{k!}\int_{-\infty}^\infty \mathrm{d}x \frac{\partial^{2k}}{\partial a^{2k}}e^{-ax^2}\Big |_{a=1} = \sum_{k=0}^\infty \frac{(-\lambda)^k}{k!} \frac{\partial^{2k}}{\partial a^{2k}}\sqrt{\frac{\pi}{a}}\Big |_{a=1} 
\label{eq:21}
\end{equation}
Of course we can easily evaluate the derivatives:
\begin{equation}
\label{eq:22}
\frac{\partial^{2k}}{\partial a^{2k}}a^{-\frac{1}{2}}\Big |_{a=1} = \underbrace{\frac{1}{2}\frac{3}{2}\cdot\frac{5}{2}\frac{7}{2}\cdot\frac{9}{2}\frac{11}{2}\cdot\,\cdots}_{\text{total of $2k$ factors}}
\end{equation}
In order to find an explicit expression for (\ref{eq:22}) one can insert factors of $1$ between all factors, such that the nominator becomes $(4k)!$:
\begin{align}
\label{eq:23}
\frac{\partial^{2k}}{\partial a^{2k}}a^{-\frac{1}{2}}\Big |_{a=1} &= \underbrace{\frac{1}{2}\frac{2}{2}\frac{3}{2}\frac{4}{4}\frac{5}{2}\frac{6}{6}\frac{7}{2}\frac{8}{8}\frac{9}{2}\frac{10}{10}\frac{11}{2}\frac{12}{12}\dots}_{\text{total of $4k$ factors}} = \frac{(4k)!}{2^{2k}}\underbrace{\frac{1}{2}\frac{1}{4}\frac{1}{6}\frac{1}{8}\frac{1}{10}\frac{1}{12}\dots}_{\text{total of $2k$ factors}} \nonumber \\
&= \frac{(4k)!}{2^{2k}}\frac{1}{2^{2k}(2k)!} = \frac{(4k)!}{2^{4k}(2k)!}
\end{align}
Thus we obtain the ``perturbative solution'' of problem (\ref{eq:16})
\begin{equation}
\label{eq:24}
\mathcal Z (\lambda) = \sum_{k=0}^\infty\sqrt{\pi}\frac{(-\lambda)^k(4k)!}{2^{4k}(2k)!k!}.
\end{equation}
Let us take a closer look at this expression.  By observing that the
denominator of the summand eventually contains smaller factors than
the nominator for all $k$ larger than a critical integer, we can
realize that the series is divergent. More carefully we can apply
Stirling's formula $n! \approx \sqrt{2\pi
  n}\left(\frac{n}{e}\right)^n$ for large values of $k$:
\begin{equation}
\label{eq:25}
\frac{(4k)!}{2^{4k}(2k)!k!} \approx \frac{4^k}{\sqrt{\pi k}}\left(\frac{k}{e}\right)^k\approx \frac{1}{\sqrt{2}\pi}4^k k!
\end{equation}
We already know that the sum
\begin{equation}
\label{eq:26}
\sum_{k=0}^\infty (-4\lambda)^k k!
\end{equation}
will diverge. This shows that the power series (\ref{eq:24}) is
divergent and in particular it is not the finite number that we are
looking for as an expression for (\ref{eq:16}).
\subsubsection{The perturbative and the full solution compared}
\begin{figure}[h]
  \centering
\includegraphics{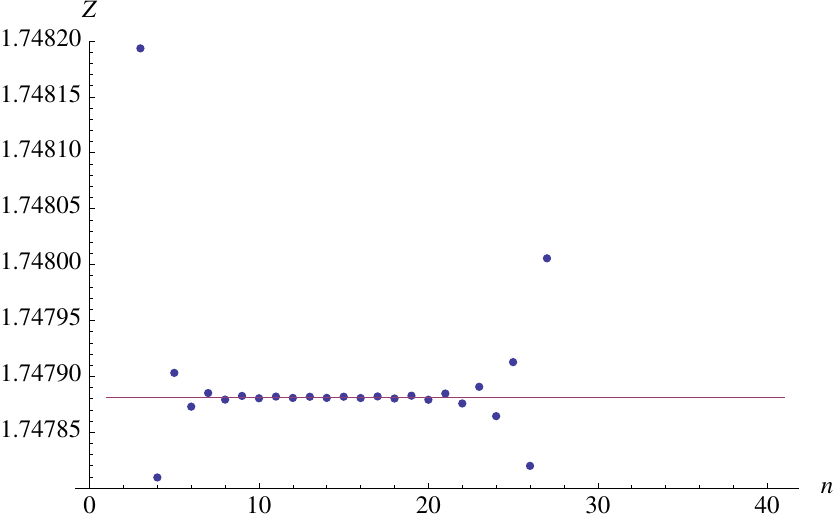}
  \caption{Values of the perturbative series (\ref{eq:24}) evaluated to order $N$}
  \label{fig:series}
\end{figure}

Even though the perturbative series will diverge, we want to
study its numerical usefulness at finite order. After all, one usually
computes only a finite number of Feynman diagrams to obtain only the
first few summands of the perturbative expansion. Is there a way to
approximate the full, non-perturbative solution (\ref{eq:18}) from
(\ref{eq:24})? Let us choose one value for $\lambda$,
e.g. $\frac{1}{50}$, and evaluate (\ref{eq:18}) numerically:
\begin{equation}
\label{eq:27}
\mathcal Z\left(\frac{1}{50}\right) = 1.7478812\dots
\end{equation}
For the same value of $\lambda$ the evaluation of the first few terms
of the infinite sum (\ref{eq:24}) 
\begin{equation}
\label{eq:24finite}
\mathcal Z_N (\lambda) = \sum_{k=0}^N\sqrt{\pi}\frac{(-\lambda)^k(4k)!}{2^{4k}(2k)!k!}.
\end{equation}
gives 
\begin{subequations}
\begin{align}
\mathcal Z_5 &= 1.7478728\dots \\
\mathcal Z_{10} &= 1.7478818\dots 
\end{align}
\end{subequations}
The first terms of the perturbative solution agree up to six digits! We can use conveniently a figure for plotting higher orders of the perturbative series. Figure \ref{fig:series} shows that the perturbative solution gets in a certain regime very close to the result of the full solution, before the series starts to diverge.
\begin{figure}[h]
  \centering
\includegraphics{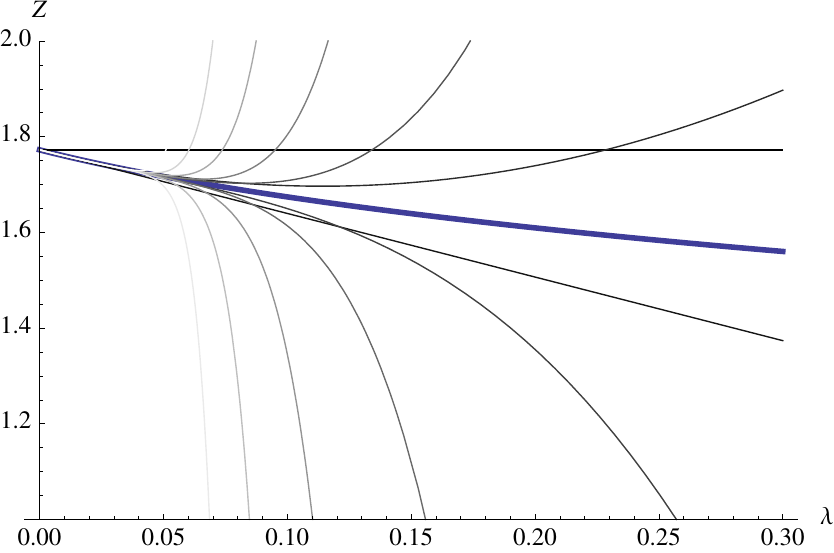}
  \caption{$\mathcal Z(\lambda)$ obtained from the full solution (thick) and first approximations from the perturbative series}
  \label{fig:serieslambda}
\end{figure}
We can use a figure as well to compare the solutions for variable $\lambda$. Figure \ref{fig:serieslambda} shows nicely the  non-perturbative solution  and compares it to the perturbative solution for orders of one to twelve. We can see that at some point all approximations given by the perturbative solution will disagree strongly from the full solution!

The question that arises is, how long does the perturbative solution
become better before it starts to diverge? Obviously, the fact that it
approximates the non-perturbative solution to high precision leads to
the great success of quantum field theory, even if for higher orders
the series diverges! As we will see now, the perturbative solution
(\ref{eq:24}) is a good approximation as long as we only consider
terms up to 
order $N =O(\frac{1}{\lambda})$. Remembering the dimensionless
coupling strength of Quantum Electrodynamics being the Sommerfeld
finestructure constant $\alpha\approx \frac{1}{137}$ we can be ensured
that perturbation theory will lead to great precision given that the
most elaborate QED calculations for $(g-2)$ are to order $N=7$!
\subsubsection{The method of steepest descend}
But what is the origin of the eventual divergence and complete loss of
numerical accuracy? It turns out that there are ``non-perturbative''
terms that do not show up in a Taylor expansion but that become
dominant when the perturbative expansion breaks down. To see this, let
us substitute $x^2\equiv \frac{u^2}{\lambda}$ in equation
(\ref{eq:16}):
\begin{equation}
\mathcal Z(\lambda)=\frac{1}{\sqrt{\lambda}}\int\mathrm{d}u e^{-\frac{u^2+u^4}{\lambda}}
\label{eq:28}
\end{equation}
The exponent is strictly negative and its absolute value becomes very large in the limit of small $\lambda$. This allows to perform the method of steepest descent: The main contribution to the integral, as $\lambda \to 0$ comes from the extrema of the integrand
\begin{equation}
u^2+u^4.
\label{eq:29}
\end{equation}
In general the method works as follow: For $\Lambda\to\infty$ we want to solve an integral of the general form
\begin{equation}
\int\mathrm{d}x A(x) e^{(i)\phi(x)\Lambda}.
\end{equation}
One expands now around its extrema\footnote{Notice that in field theory $\phi^\prime(x)=0$ is the equation of motion} $\phi^\prime(x_0)=0$ and obtains again an integral of ``Gaussian times polynomial'' type\footnote{Corrections from $(x-x_0)A^\prime(x_0)$ can be obtained by doing again the trick of smuggling an $a$ into the exponent and write the term as derivative with respect of $a$ evaluated at $a=1$.}
\begin{align}
&=\sum_{x_0:\phi'(x_0)=0}\int\mathrm{d}x(A(x_0)+(x-x_0)A^\prime(x_0)\dots)e^{\Lambda(\phi(x_0)+(x-x_0)^2\phi^{\prime\prime}(x_0)+\dots)} \nonumber \\
&= \sum_{x_0:\phi'(x_0)=0}A(x_0)e^{\Lambda\phi(x_0)}\sqrt{\frac{2\pi}{\phi^{\prime\prime}(x_0)\Lambda}}\left(1+ O \left(\frac{1}{\Lambda}\right)\right).
\label{eq:30}
\end{align}
In our case, the extrema of (\ref{eq:29}) are $u=0$ and $u=\pm i/\sqrt
2$. Expansion around the first yields the perturbative expansion of
above. The other two yield contributions like $e^{\frac 1{4\lambda}}$
that are invisible to a Taylor expansion around $\lambda=0$, as all
derivative vanish here. We have found an example of a
``non-perturbative contribution''.

The perturbative solution, however, gives meaningful results, as long
as its terms are bigger than to the non-perturbative
contributions. This allows an estimate, to what order in the
perturbative series the expansion around $\lambda=0$ dominates. This
happens also to be the order at which the divergence from the exat
solution starts as we are missing the non-perturbative terms:
\begin{align}
e^{-\frac{1}{4\lambda}} &\approx \lambda^k \nonumber \\
-\frac{1}{4\lambda} &\approx k\cdot\ln(\lambda) \nonumber \\
k &\approx \frac{1}{\lambda}
\end{align}

We have seen that the perturbative analysis of (\ref{eq:28}) requires
as well expansions around the other extrema, besides $\lambda = 0$!
Combining all power series together, the resulting perturbative
solution has a chance to converge. Before we continue to the
mathematical discussion of the problem how finite results can be
obtained from divergent series, we will take a look on examples of
non-perturbative contributions in physics.

\subsubsection{Instantons}
``Field configurations'' contributing $e^{-\frac{1}{\sim\lambda}}$ are
called ``instantons''. Usually these contributions are hard to
calculate, in some situations, however, one can find the
result. Consider for example a gauge theory\footnote{Hodge $\ast$
  operator: $\ast
  F^{\mu\nu}\equiv\epsilon^{\mu\nu\sigma\tau}F_{\sigma\tau}$. More
  details can be found in chapters 1.10 and 10.5 of \cite{Nakahara}.}
\begin{equation}
S = \int\mathcal L = \int \frac{1}{g^2}\tr(F\wedge \ast F)
\label{eq:gauge}
\end{equation}
The stationary point we use for the expansion is given by the equations of motion
\begin{subequations}
\begin{align}
\mathrm d F &= 0  \label{eq:eom1}\\
\mathrm d \ast F &= 0 \label{eq:eom2}
\end{align}
\end{subequations}
The first equation (\ref{eq:eom1}) is automatically fulfilled once we
express the field-strength in terms of a vector potential $F = \mathrm
d A$. The second (\ref{eq:eom2}) is automatically solved if it happens that
\begin{equation}
F = \ast F.
\label{eq:instanton}
\end{equation}
One calls solutions to (\ref{eq:instanton}) instantons. In terms of
the vector potential $A$ (\ref{eq:instanton}) is a first order partial
differential equation as compared to (\ref{eq:eom2}) which is second order. One can easily see that there exist no solution
in Lorentzian metric as the Hodge star squares to $-1$ on 2-forms.
\begin{equation}
F = \ast F = \ast\ast F = -F.
\label{eq:Lorentz}
\end{equation}
In Euclidean metric, however, such solutions exist because of
$\ast\ast F = F$. As it turns out (as one can for example argue using
the Atiya-Singer index theorem), for a compact manifold $M$, the
action in the instanton case yields an integer (up to a pre-factor):
\begin{equation}
\int_M \tr(F \wedge F) \in 8\pi^2 \MZ
\label{eq:index}
\end{equation}
This leads to
\begin{equation}
e^{\frac{1}{g^2}\int \tr(F \wedge \ast F)} = e^{\frac{1}{g^2}\int \tr(F \wedge F)} = e^{-\frac{1}{g^2}8\pi^2 N}  
\label{eq:result}
\end{equation}

\subsubsection{Dual theories}
Sometimes a quantum field theory with coupling constant $\lambda$ can
be rewritten in terms of another (or possibly the same) quantum field
theory with coupling $\tilde\lambda=1/{\lambda}$. One calls such a
relation between two theories a ``duality''. In many examples, such
theories arise from string theory constructions, where the coupling
$\lambda$ can be given a geometric meaning. Imagine for example a
problem of a quantum field theory on a torus. A torus can be viewed as
$\MC / (\MZ + \tau\MZ)$ with $\tau \in \MC \backslash \MR$. The torus
has a basis of two non-contractible circles, one that goes along the
real axis from 0 to 1 and one that goes from 0 to $\tau$. This choice
of basis, however, is not unique: For example, swapping these cycles
corresponds to a substitution $\tau \rightarrow -1/{\tau}$. If the
torus parameter $\tau$ is identified with the coupling strength a
duality has been found since both $\tau$ and $-1/\tau$ describe
geometricaly the same torus! To make contact with our discussion
above, we should identify $\lambda$ with the imaginary part of
$\tau$. The duality allows for a Taylor expansion of the
non-perturbative contributions via
\begin{equation}
e^{-\frac{1}{4\lambda}}=e^{-\frac{\tilde\lambda}{4}}=\sum_k^\infty \frac{\left(-\tilde\lambda\right)^k}{k!4^k}
\label{eq:dual}
\end{equation}
Troublesome terms in one theory are therefore perfectly defined in the
dual theory. The caveat however is the difficulty of actually proving
that $\lambda\rightarrow 1/{\lambda}$ is a symmetry of the
quantum field theory at hand.

\subsection{Asymptotic series and Borel summation}
In the following, we take a look on the mathematical situation of asymptotic series. This discussion is based on chapter $XII$ of \cite{ReedSimon}.
\begin{definition}
Let $f\colon\MR_{\geq 0} \to \MC$. The series $\sum_n^\infty a_n z^n$
is called \textnormal{asymptotic} to $f$ as $z\searrow 0$ iff
\begin{equation}
\forall N \in \MN: \lim_{z\searrow 0} \frac{f(z)-\sum_n^N a_n z^n}{z^n} = 0
\label{eq:def1}
\end{equation}
For $z \in \MC$ a analog definition is possible. 
\label{def:asymptotic}
\end{definition}
Obviously, every function can have at most one asymptotic
expansion. This can be seen by assuming two asymptotic expansions
$a_n$ and $\tilde a_n$. (\ref{eq:def1}) requires that $a_n = \tilde
a_n$. Otherwise, let $n$ be the smallest index for which $a_n\ne
\tilde a_n$ and 
\begin{equation}
\lim_{z\searrow 0}\frac{\sum_k(a_k - \tilde a_k)z^k}{z^n} = a_n - \tilde a_n \stackrel{!}{=} 0.
\label{eq:expansion}
\end{equation}
The other way around is not true, as can be seen by $f(z) = e^{-\frac{1}{z}}$ and $\tilde f (z) = 0$ having both the asymptotic series $\sum_k^\infty 0\cdot z^k$. This means that knowing the asymptotic series of a function tells us nothing about $f(z)$ for a non vanishing $z$, we only know how $f(z)$ approaches $f(0)$ as $z \searrow 0$.

We try to find a stronger definition of an asymptotic series, allowing us to uniquely recover one function. The following theorem helps us to find the necessary conditon:
\begin{theorem}
  (\textbf{Carleman's theorem}) Let $g$ be an analytic function in the
  interior of $S = \{z\in\MC|\abs{z}\leq B,\abs{\mathrm{arg}\ z}\leq
  \frac{\pi}{2}\}$ and continous on $S$. If for all $n \in \MN$ and $z
  \in S$ we have $\abs{g(z)}~\leq~b_n\abs{z}^n$ and $\sum_n^\infty
  b_n^{-\frac{1}{n}}=\infty$, then $g$ is identically
  zero.

A simpler special case of the theorem is found by considering $g$ an analytic function in the interior of $S_\epsilon = \{z\in\MC|\left|z\right|\leq R,\abs{\mathrm{arg}\ z}\leq \frac{\pi}{2}+\epsilon\}$ for some $\epsilon > 0$ and continous on $S_\epsilon$. If there exist $C$ and $B$ so that $\abs{g(z)}<CB^nn!\abs{z}^n\ \forall z\in S\ and\ \forall n$, then $g$ is identically zero.
\label{Carleman}
\end{theorem}
In order to find a unique function for an asymptotic series, we use Carleman's theorem to define ``strong asymptotic series''.
\begin{definition}
Let $f$ be an analytic function on the interior of $S_\epsilon = \{z\in\MC|\abs{z}\leq R,\abs{\mathrm{arg}\ z}\leq \frac{\pi}{2}+\epsilon\} \rightarrow \MR$. The series $\sum_n^\infty a_n z^n$ is a \textnormal{strong asymptotic series} if there exist $C,\sigma$ so that\ $\forall N\in\MN,z\in S_\epsilon$ the \textnormal{strong asymptotic condition}
\begin{equation}
\abs{f(z)-\sum_n^N a_n z^n}\leq C\sigma^{N+1}(N+1)!\abs{z}^{N+1}
\label{eq:condition}
\end{equation}
is fulfilled.
\end{definition}
This means, if we are given a strong asymptotic series, we can recover by theorem \ref{Carleman} the function! Assume for example $\sum_n^\infty a_n z^n$ is a strong asymptotic series for two functions $f$ and $g$, respectively. Then
\begin{equation}
\abs{f(z)-g(z)}\leq 2C\sigma^{N+1}(N+1)!\abs{z}^{N+1} \Rightarrow f=g
\label{eq:f=g}
\end{equation}
The strong asymptotic condition (\ref{eq:condition}) implies
$\abs{a_n} \leq C\sigma^n n!$. This is precisely the growth behavior
of (\ref{eq:24}) we found in our toy example, where
$C=\frac{1}{\sqrt{2\pi}}$ and $\sigma=4$. The necessary conditions,
therefore, are fulfilled in our toy model (assuming analyticity away
from 0 of course).

By now, we learned that a strong asymptotic series (in particular the
type we obtain in quantum field theory) although not converging has
the chance to be  a unique approximation to one function. The final
question is, how one can obtain this function $f$ from its strong
asymptotic series. In the last theorem we introduce the method of
``Borel summation'' to obtain a final result. We can define a
convergent series by taking out a factor of $n!$ from the coefficients:
\begin{theorem}
(\textbf{Watson's theorem})
If $f:S_\epsilon\to\MR$ has a strong asymptotic series $\sum_n^\infty a_n z^n$, we define the \textnormal{Borel transform}
\begin{equation}
g(z) = \sum_n^\infty\frac{a_n}{n!}z^n.
\label{eq:transform}
\end{equation}
The Borel transform converges for $\abs{z}<\frac{1}{\abs{\sigma}}$. We
obtained a convergent power series with finite radius of convergence,
which, as it turns out, can be analytically continued to all complex
$z\in\MC$ with $\abs{\mathrm{arg}\ z}<\epsilon$. Then the function $f$
is given by the Laplace transform
\begin{equation}
f(z)=\int_0^\infty\mathrm{d}b\ g(bz)e^{-b}.
\label{eq:functionf}
\end{equation}
\end{theorem}
This Laplace transform is called ``inverse Borel transform" and the method outlined here is known as ``Borel summability method''. It describes how to obtain a finite answer from divergent series, that is formally a sum for the series.

Let us make a sanity check. Using $\int_0^\infty \mathrm{d}x\ x^k e^{-x}=k!$ we can plug the definition of the Borel transform into (\ref{eq:functionf}), formally interchange the sum and the integration and obtain
\begin{equation}
f(z)=\int_0^\infty\mathrm{d}b\ g(bz)e^{-b} = \int_0^\infty\mathrm{d}b\sum_n\frac{a_n}{n!}b^nz^ne^{-b}\ \hbox{``=''} \sum_n a_n z^n.
\label{eq:final}
\end{equation}
So at least for analytic functions we do recover the original function.
\subsection{Summary}
We have learned why for $N<\mathcal O (\frac{1}{\lambda})$ the sum of
the first $N$ terms of the perturbation expansion is numerically good,
even when the original series $\sum_n a_n z^n = \infty$ diverges. This
way we approximate the true function up to instantonic terms of the
order of $e^{1/\lambda}$ which a Taylor expansion cannot
resolve. Given that the coefficients $a_n$ obey the strong asymptotic
condition $\left|a_n\right|\leq C\sigma^nn!$, which is usually the
case when using Feynman diagrams, the Borel transform exists and one
can compute the Borel summation. Unfortunately this is more a
theoretical assurance that perturbation theory can be given a
mathematical meaning even though it does not converge, since in order
to really compute the integral (\ref{eq:functionf}) one has to know
the analytic continuation of $g$ which requires knowledge of
\textit{all} coefficients $a_n$ and not just the first $N$.

%% file: potential1.tex
\setlength{\unitlength}{0.240900pt}
\ifx\plotpoint\undefined\newsavebox{\plotpoint}\fi
\begin{picture}(826,590)(0,0)
\sbox{\plotpoint}{\rule[-0.200pt]{0.400pt}{0.400pt}}%
\put(50.0,295.0){\rule[-0.200pt]{172.243pt}{0.400pt}}
\put(408.0,40.0){\rule[-0.200pt]{0.400pt}{122.618pt}}
\put(434,549){\makebox(0,0)[l]{$V(\phi)$}}
\put(765,335){\makebox(0,0)[l]{$\phi$}}
\multiput(109.59,542.77)(0.482,-1.847){9}{\rule{0.116pt}{1.500pt}}
\multiput(108.17,545.89)(6.000,-17.887){2}{\rule{0.400pt}{0.750pt}}
\multiput(115.59,522.37)(0.485,-1.637){11}{\rule{0.117pt}{1.357pt}}
\multiput(114.17,525.18)(7.000,-19.183){2}{\rule{0.400pt}{0.679pt}}
\multiput(122.59,500.60)(0.485,-1.560){11}{\rule{0.117pt}{1.300pt}}
\multiput(121.17,503.30)(7.000,-18.302){2}{\rule{0.400pt}{0.650pt}}
\multiput(129.59,480.64)(0.488,-1.220){13}{\rule{0.117pt}{1.050pt}}
\multiput(128.17,482.82)(8.000,-16.821){2}{\rule{0.400pt}{0.525pt}}
\multiput(137.59,461.55)(0.485,-1.255){11}{\rule{0.117pt}{1.071pt}}
\multiput(136.17,463.78)(7.000,-14.776){2}{\rule{0.400pt}{0.536pt}}
\multiput(144.59,444.55)(0.485,-1.255){11}{\rule{0.117pt}{1.071pt}}
\multiput(143.17,446.78)(7.000,-14.776){2}{\rule{0.400pt}{0.536pt}}
\multiput(151.59,428.03)(0.485,-1.103){11}{\rule{0.117pt}{0.957pt}}
\multiput(150.17,430.01)(7.000,-13.013){2}{\rule{0.400pt}{0.479pt}}
\multiput(158.59,413.89)(0.488,-0.824){13}{\rule{0.117pt}{0.750pt}}
\multiput(157.17,415.44)(8.000,-11.443){2}{\rule{0.400pt}{0.375pt}}
\multiput(166.59,400.50)(0.485,-0.950){11}{\rule{0.117pt}{0.843pt}}
\multiput(165.17,402.25)(7.000,-11.251){2}{\rule{0.400pt}{0.421pt}}
\multiput(173.59,387.98)(0.485,-0.798){11}{\rule{0.117pt}{0.729pt}}
\multiput(172.17,389.49)(7.000,-9.488){2}{\rule{0.400pt}{0.364pt}}
\multiput(180.59,377.21)(0.485,-0.721){11}{\rule{0.117pt}{0.671pt}}
\multiput(179.17,378.61)(7.000,-8.606){2}{\rule{0.400pt}{0.336pt}}
\multiput(187.59,367.21)(0.485,-0.721){11}{\rule{0.117pt}{0.671pt}}
\multiput(186.17,368.61)(7.000,-8.606){2}{\rule{0.400pt}{0.336pt}}
\multiput(194.00,358.93)(0.494,-0.488){13}{\rule{0.500pt}{0.117pt}}
\multiput(194.00,359.17)(6.962,-8.000){2}{\rule{0.250pt}{0.400pt}}
\multiput(202.59,349.69)(0.485,-0.569){11}{\rule{0.117pt}{0.557pt}}
\multiput(201.17,350.84)(7.000,-6.844){2}{\rule{0.400pt}{0.279pt}}
\multiput(209.00,342.93)(0.492,-0.485){11}{\rule{0.500pt}{0.117pt}}
\multiput(209.00,343.17)(5.962,-7.000){2}{\rule{0.250pt}{0.400pt}}
\multiput(216.00,335.93)(0.581,-0.482){9}{\rule{0.567pt}{0.116pt}}
\multiput(216.00,336.17)(5.824,-6.000){2}{\rule{0.283pt}{0.400pt}}
\multiput(223.00,329.93)(0.821,-0.477){7}{\rule{0.740pt}{0.115pt}}
\multiput(223.00,330.17)(6.464,-5.000){2}{\rule{0.370pt}{0.400pt}}
\multiput(231.00,324.93)(0.710,-0.477){7}{\rule{0.660pt}{0.115pt}}
\multiput(231.00,325.17)(5.630,-5.000){2}{\rule{0.330pt}{0.400pt}}
\multiput(238.00,319.94)(0.920,-0.468){5}{\rule{0.800pt}{0.113pt}}
\multiput(238.00,320.17)(5.340,-4.000){2}{\rule{0.400pt}{0.400pt}}
\multiput(245.00,315.94)(0.920,-0.468){5}{\rule{0.800pt}{0.113pt}}
\multiput(245.00,316.17)(5.340,-4.000){2}{\rule{0.400pt}{0.400pt}}
\multiput(252.00,311.95)(1.355,-0.447){3}{\rule{1.033pt}{0.108pt}}
\multiput(252.00,312.17)(4.855,-3.000){2}{\rule{0.517pt}{0.400pt}}
\multiput(259.00,308.95)(1.579,-0.447){3}{\rule{1.167pt}{0.108pt}}
\multiput(259.00,309.17)(5.579,-3.000){2}{\rule{0.583pt}{0.400pt}}
\put(267,305.17){\rule{1.500pt}{0.400pt}}
\multiput(267.00,306.17)(3.887,-2.000){2}{\rule{0.750pt}{0.400pt}}
\put(274,303.17){\rule{1.500pt}{0.400pt}}
\multiput(274.00,304.17)(3.887,-2.000){2}{\rule{0.750pt}{0.400pt}}
\put(281,301.17){\rule{1.500pt}{0.400pt}}
\multiput(281.00,302.17)(3.887,-2.000){2}{\rule{0.750pt}{0.400pt}}
\put(288,299.67){\rule{1.927pt}{0.400pt}}
\multiput(288.00,300.17)(4.000,-1.000){2}{\rule{0.964pt}{0.400pt}}
\put(296,298.17){\rule{1.500pt}{0.400pt}}
\multiput(296.00,299.17)(3.887,-2.000){2}{\rule{0.750pt}{0.400pt}}
\put(303,296.67){\rule{1.686pt}{0.400pt}}
\multiput(303.00,297.17)(3.500,-1.000){2}{\rule{0.843pt}{0.400pt}}
\put(317,295.67){\rule{1.686pt}{0.400pt}}
\multiput(317.00,296.17)(3.500,-1.000){2}{\rule{0.843pt}{0.400pt}}
\put(310.0,297.0){\rule[-0.200pt]{1.686pt}{0.400pt}}
\put(332,294.67){\rule{1.686pt}{0.400pt}}
\multiput(332.00,295.17)(3.500,-1.000){2}{\rule{0.843pt}{0.400pt}}
\put(324.0,296.0){\rule[-0.200pt]{1.927pt}{0.400pt}}
\put(476,294.67){\rule{1.686pt}{0.400pt}}
\multiput(476.00,294.17)(3.500,1.000){2}{\rule{0.843pt}{0.400pt}}
\put(339.0,295.0){\rule[-0.200pt]{33.003pt}{0.400pt}}
\put(491,295.67){\rule{1.686pt}{0.400pt}}
\multiput(491.00,295.17)(3.500,1.000){2}{\rule{0.843pt}{0.400pt}}
\put(483.0,296.0){\rule[-0.200pt]{1.927pt}{0.400pt}}
\put(505,296.67){\rule{1.686pt}{0.400pt}}
\multiput(505.00,296.17)(3.500,1.000){2}{\rule{0.843pt}{0.400pt}}
\put(512,298.17){\rule{1.500pt}{0.400pt}}
\multiput(512.00,297.17)(3.887,2.000){2}{\rule{0.750pt}{0.400pt}}
\put(519,299.67){\rule{1.927pt}{0.400pt}}
\multiput(519.00,299.17)(4.000,1.000){2}{\rule{0.964pt}{0.400pt}}
\put(527,301.17){\rule{1.500pt}{0.400pt}}
\multiput(527.00,300.17)(3.887,2.000){2}{\rule{0.750pt}{0.400pt}}
\put(534,303.17){\rule{1.500pt}{0.400pt}}
\multiput(534.00,302.17)(3.887,2.000){2}{\rule{0.750pt}{0.400pt}}
\put(541,305.17){\rule{1.500pt}{0.400pt}}
\multiput(541.00,304.17)(3.887,2.000){2}{\rule{0.750pt}{0.400pt}}
\multiput(548.00,307.61)(1.579,0.447){3}{\rule{1.167pt}{0.108pt}}
\multiput(548.00,306.17)(5.579,3.000){2}{\rule{0.583pt}{0.400pt}}
\multiput(556.00,310.61)(1.355,0.447){3}{\rule{1.033pt}{0.108pt}}
\multiput(556.00,309.17)(4.855,3.000){2}{\rule{0.517pt}{0.400pt}}
\multiput(563.00,313.60)(0.920,0.468){5}{\rule{0.800pt}{0.113pt}}
\multiput(563.00,312.17)(5.340,4.000){2}{\rule{0.400pt}{0.400pt}}
\multiput(570.00,317.60)(0.920,0.468){5}{\rule{0.800pt}{0.113pt}}
\multiput(570.00,316.17)(5.340,4.000){2}{\rule{0.400pt}{0.400pt}}
\multiput(577.00,321.59)(0.710,0.477){7}{\rule{0.660pt}{0.115pt}}
\multiput(577.00,320.17)(5.630,5.000){2}{\rule{0.330pt}{0.400pt}}
\multiput(584.00,326.59)(0.821,0.477){7}{\rule{0.740pt}{0.115pt}}
\multiput(584.00,325.17)(6.464,5.000){2}{\rule{0.370pt}{0.400pt}}
\multiput(592.00,331.59)(0.581,0.482){9}{\rule{0.567pt}{0.116pt}}
\multiput(592.00,330.17)(5.824,6.000){2}{\rule{0.283pt}{0.400pt}}
\multiput(599.00,337.59)(0.492,0.485){11}{\rule{0.500pt}{0.117pt}}
\multiput(599.00,336.17)(5.962,7.000){2}{\rule{0.250pt}{0.400pt}}
\multiput(606.59,344.00)(0.485,0.569){11}{\rule{0.117pt}{0.557pt}}
\multiput(605.17,344.00)(7.000,6.844){2}{\rule{0.400pt}{0.279pt}}
\multiput(613.00,352.59)(0.494,0.488){13}{\rule{0.500pt}{0.117pt}}
\multiput(613.00,351.17)(6.962,8.000){2}{\rule{0.250pt}{0.400pt}}
\multiput(621.59,360.00)(0.485,0.721){11}{\rule{0.117pt}{0.671pt}}
\multiput(620.17,360.00)(7.000,8.606){2}{\rule{0.400pt}{0.336pt}}
\multiput(628.59,370.00)(0.485,0.721){11}{\rule{0.117pt}{0.671pt}}
\multiput(627.17,370.00)(7.000,8.606){2}{\rule{0.400pt}{0.336pt}}
\multiput(635.59,380.00)(0.485,0.798){11}{\rule{0.117pt}{0.729pt}}
\multiput(634.17,380.00)(7.000,9.488){2}{\rule{0.400pt}{0.364pt}}
\multiput(642.59,391.00)(0.485,0.950){11}{\rule{0.117pt}{0.843pt}}
\multiput(641.17,391.00)(7.000,11.251){2}{\rule{0.400pt}{0.421pt}}
\multiput(649.59,404.00)(0.488,0.824){13}{\rule{0.117pt}{0.750pt}}
\multiput(648.17,404.00)(8.000,11.443){2}{\rule{0.400pt}{0.375pt}}
\multiput(657.59,417.00)(0.485,1.103){11}{\rule{0.117pt}{0.957pt}}
\multiput(656.17,417.00)(7.000,13.013){2}{\rule{0.400pt}{0.479pt}}
\multiput(664.59,432.00)(0.485,1.255){11}{\rule{0.117pt}{1.071pt}}
\multiput(663.17,432.00)(7.000,14.776){2}{\rule{0.400pt}{0.536pt}}
\multiput(671.59,449.00)(0.485,1.255){11}{\rule{0.117pt}{1.071pt}}
\multiput(670.17,449.00)(7.000,14.776){2}{\rule{0.400pt}{0.536pt}}
\multiput(678.59,466.00)(0.488,1.220){13}{\rule{0.117pt}{1.050pt}}
\multiput(677.17,466.00)(8.000,16.821){2}{\rule{0.400pt}{0.525pt}}
\multiput(686.59,485.00)(0.485,1.560){11}{\rule{0.117pt}{1.300pt}}
\multiput(685.17,485.00)(7.000,18.302){2}{\rule{0.400pt}{0.650pt}}
\multiput(693.59,506.00)(0.485,1.637){11}{\rule{0.117pt}{1.357pt}}
\multiput(692.17,506.00)(7.000,19.183){2}{\rule{0.400pt}{0.679pt}}
\multiput(700.59,528.00)(0.482,1.847){9}{\rule{0.116pt}{1.500pt}}
\multiput(699.17,528.00)(6.000,17.887){2}{\rule{0.400pt}{0.750pt}}
\put(498.0,297.0){\rule[-0.200pt]{1.686pt}{0.400pt}}
\end{picture}

%% file: potential2.tex
\setlength{\unitlength}{0.240900pt}
\ifx\plotpoint\undefined\newsavebox{\plotpoint}\fi
\begin{picture}(826,590)(0,0)
\sbox{\plotpoint}{\rule[-0.200pt]{0.400pt}{0.400pt}}%
\put(50.0,295.0){\rule[-0.200pt]{172.243pt}{0.400pt}}
\put(408.0,40.0){\rule[-0.200pt]{0.400pt}{122.618pt}}
\put(434,549){\makebox(0,0)[l]{$V(\phi)$}}
\put(765,335){\makebox(0,0)[l]{$\phi$}}
\multiput(109.59,40.00)(0.482,1.847){9}{\rule{0.116pt}{1.500pt}}
\multiput(108.17,40.00)(6.000,17.887){2}{\rule{0.400pt}{0.750pt}}
\multiput(115.59,61.00)(0.485,1.637){11}{\rule{0.117pt}{1.357pt}}
\multiput(114.17,61.00)(7.000,19.183){2}{\rule{0.400pt}{0.679pt}}
\multiput(122.59,83.00)(0.485,1.560){11}{\rule{0.117pt}{1.300pt}}
\multiput(121.17,83.00)(7.000,18.302){2}{\rule{0.400pt}{0.650pt}}
\multiput(129.59,104.00)(0.488,1.220){13}{\rule{0.117pt}{1.050pt}}
\multiput(128.17,104.00)(8.000,16.821){2}{\rule{0.400pt}{0.525pt}}
\multiput(137.59,123.00)(0.485,1.255){11}{\rule{0.117pt}{1.071pt}}
\multiput(136.17,123.00)(7.000,14.776){2}{\rule{0.400pt}{0.536pt}}
\multiput(144.59,140.00)(0.485,1.255){11}{\rule{0.117pt}{1.071pt}}
\multiput(143.17,140.00)(7.000,14.776){2}{\rule{0.400pt}{0.536pt}}
\multiput(151.59,157.00)(0.485,1.103){11}{\rule{0.117pt}{0.957pt}}
\multiput(150.17,157.00)(7.000,13.013){2}{\rule{0.400pt}{0.479pt}}
\multiput(158.59,172.00)(0.488,0.824){13}{\rule{0.117pt}{0.750pt}}
\multiput(157.17,172.00)(8.000,11.443){2}{\rule{0.400pt}{0.375pt}}
\multiput(166.59,185.00)(0.485,0.950){11}{\rule{0.117pt}{0.843pt}}
\multiput(165.17,185.00)(7.000,11.251){2}{\rule{0.400pt}{0.421pt}}
\multiput(173.59,198.00)(0.485,0.798){11}{\rule{0.117pt}{0.729pt}}
\multiput(172.17,198.00)(7.000,9.488){2}{\rule{0.400pt}{0.364pt}}
\multiput(180.59,209.00)(0.485,0.721){11}{\rule{0.117pt}{0.671pt}}
\multiput(179.17,209.00)(7.000,8.606){2}{\rule{0.400pt}{0.336pt}}
\multiput(187.59,219.00)(0.485,0.721){11}{\rule{0.117pt}{0.671pt}}
\multiput(186.17,219.00)(7.000,8.606){2}{\rule{0.400pt}{0.336pt}}
\multiput(194.00,229.59)(0.494,0.488){13}{\rule{0.500pt}{0.117pt}}
\multiput(194.00,228.17)(6.962,8.000){2}{\rule{0.250pt}{0.400pt}}
\multiput(202.59,237.00)(0.485,0.569){11}{\rule{0.117pt}{0.557pt}}
\multiput(201.17,237.00)(7.000,6.844){2}{\rule{0.400pt}{0.279pt}}
\multiput(209.00,245.59)(0.492,0.485){11}{\rule{0.500pt}{0.117pt}}
\multiput(209.00,244.17)(5.962,7.000){2}{\rule{0.250pt}{0.400pt}}
\multiput(216.00,252.59)(0.581,0.482){9}{\rule{0.567pt}{0.116pt}}
\multiput(216.00,251.17)(5.824,6.000){2}{\rule{0.283pt}{0.400pt}}
\multiput(223.00,258.59)(0.821,0.477){7}{\rule{0.740pt}{0.115pt}}
\multiput(223.00,257.17)(6.464,5.000){2}{\rule{0.370pt}{0.400pt}}
\multiput(231.00,263.59)(0.710,0.477){7}{\rule{0.660pt}{0.115pt}}
\multiput(231.00,262.17)(5.630,5.000){2}{\rule{0.330pt}{0.400pt}}
\multiput(238.00,268.60)(0.920,0.468){5}{\rule{0.800pt}{0.113pt}}
\multiput(238.00,267.17)(5.340,4.000){2}{\rule{0.400pt}{0.400pt}}
\multiput(245.00,272.60)(0.920,0.468){5}{\rule{0.800pt}{0.113pt}}
\multiput(245.00,271.17)(5.340,4.000){2}{\rule{0.400pt}{0.400pt}}
\multiput(252.00,276.61)(1.355,0.447){3}{\rule{1.033pt}{0.108pt}}
\multiput(252.00,275.17)(4.855,3.000){2}{\rule{0.517pt}{0.400pt}}
\multiput(259.00,279.61)(1.579,0.447){3}{\rule{1.167pt}{0.108pt}}
\multiput(259.00,278.17)(5.579,3.000){2}{\rule{0.583pt}{0.400pt}}
\put(267,282.17){\rule{1.500pt}{0.400pt}}
\multiput(267.00,281.17)(3.887,2.000){2}{\rule{0.750pt}{0.400pt}}
\put(274,284.17){\rule{1.500pt}{0.400pt}}
\multiput(274.00,283.17)(3.887,2.000){2}{\rule{0.750pt}{0.400pt}}
\put(281,286.17){\rule{1.500pt}{0.400pt}}
\multiput(281.00,285.17)(3.887,2.000){2}{\rule{0.750pt}{0.400pt}}
\put(288,287.67){\rule{1.927pt}{0.400pt}}
\multiput(288.00,287.17)(4.000,1.000){2}{\rule{0.964pt}{0.400pt}}
\put(296,289.17){\rule{1.500pt}{0.400pt}}
\multiput(296.00,288.17)(3.887,2.000){2}{\rule{0.750pt}{0.400pt}}
\put(303,290.67){\rule{1.686pt}{0.400pt}}
\multiput(303.00,290.17)(3.500,1.000){2}{\rule{0.843pt}{0.400pt}}
\put(317,291.67){\rule{1.686pt}{0.400pt}}
\multiput(317.00,291.17)(3.500,1.000){2}{\rule{0.843pt}{0.400pt}}
\put(310.0,292.0){\rule[-0.200pt]{1.686pt}{0.400pt}}
\put(332,292.67){\rule{1.686pt}{0.400pt}}
\multiput(332.00,292.17)(3.500,1.000){2}{\rule{0.843pt}{0.400pt}}
\put(324.0,293.0){\rule[-0.200pt]{1.927pt}{0.400pt}}
\put(476,292.67){\rule{1.686pt}{0.400pt}}
\multiput(476.00,293.17)(3.500,-1.000){2}{\rule{0.843pt}{0.400pt}}
\put(339.0,294.0){\rule[-0.200pt]{33.003pt}{0.400pt}}
\put(491,291.67){\rule{1.686pt}{0.400pt}}
\multiput(491.00,292.17)(3.500,-1.000){2}{\rule{0.843pt}{0.400pt}}
\put(483.0,293.0){\rule[-0.200pt]{1.927pt}{0.400pt}}
\put(505,290.67){\rule{1.686pt}{0.400pt}}
\multiput(505.00,291.17)(3.500,-1.000){2}{\rule{0.843pt}{0.400pt}}
\put(512,289.17){\rule{1.500pt}{0.400pt}}
\multiput(512.00,290.17)(3.887,-2.000){2}{\rule{0.750pt}{0.400pt}}
\put(519,287.67){\rule{1.927pt}{0.400pt}}
\multiput(519.00,288.17)(4.000,-1.000){2}{\rule{0.964pt}{0.400pt}}
\put(527,286.17){\rule{1.500pt}{0.400pt}}
\multiput(527.00,287.17)(3.887,-2.000){2}{\rule{0.750pt}{0.400pt}}
\put(534,284.17){\rule{1.500pt}{0.400pt}}
\multiput(534.00,285.17)(3.887,-2.000){2}{\rule{0.750pt}{0.400pt}}
\put(541,282.17){\rule{1.500pt}{0.400pt}}
\multiput(541.00,283.17)(3.887,-2.000){2}{\rule{0.750pt}{0.400pt}}
\multiput(548.00,280.95)(1.579,-0.447){3}{\rule{1.167pt}{0.108pt}}
\multiput(548.00,281.17)(5.579,-3.000){2}{\rule{0.583pt}{0.400pt}}
\multiput(556.00,277.95)(1.355,-0.447){3}{\rule{1.033pt}{0.108pt}}
\multiput(556.00,278.17)(4.855,-3.000){2}{\rule{0.517pt}{0.400pt}}
\multiput(563.00,274.94)(0.920,-0.468){5}{\rule{0.800pt}{0.113pt}}
\multiput(563.00,275.17)(5.340,-4.000){2}{\rule{0.400pt}{0.400pt}}
\multiput(570.00,270.94)(0.920,-0.468){5}{\rule{0.800pt}{0.113pt}}
\multiput(570.00,271.17)(5.340,-4.000){2}{\rule{0.400pt}{0.400pt}}
\multiput(577.00,266.93)(0.710,-0.477){7}{\rule{0.660pt}{0.115pt}}
\multiput(577.00,267.17)(5.630,-5.000){2}{\rule{0.330pt}{0.400pt}}
\multiput(584.00,261.93)(0.821,-0.477){7}{\rule{0.740pt}{0.115pt}}
\multiput(584.00,262.17)(6.464,-5.000){2}{\rule{0.370pt}{0.400pt}}
\multiput(592.00,256.93)(0.581,-0.482){9}{\rule{0.567pt}{0.116pt}}
\multiput(592.00,257.17)(5.824,-6.000){2}{\rule{0.283pt}{0.400pt}}
\multiput(599.00,250.93)(0.492,-0.485){11}{\rule{0.500pt}{0.117pt}}
\multiput(599.00,251.17)(5.962,-7.000){2}{\rule{0.250pt}{0.400pt}}
\multiput(606.59,242.69)(0.485,-0.569){11}{\rule{0.117pt}{0.557pt}}
\multiput(605.17,243.84)(7.000,-6.844){2}{\rule{0.400pt}{0.279pt}}
\multiput(613.00,235.93)(0.494,-0.488){13}{\rule{0.500pt}{0.117pt}}
\multiput(613.00,236.17)(6.962,-8.000){2}{\rule{0.250pt}{0.400pt}}
\multiput(621.59,226.21)(0.485,-0.721){11}{\rule{0.117pt}{0.671pt}}
\multiput(620.17,227.61)(7.000,-8.606){2}{\rule{0.400pt}{0.336pt}}
\multiput(628.59,216.21)(0.485,-0.721){11}{\rule{0.117pt}{0.671pt}}
\multiput(627.17,217.61)(7.000,-8.606){2}{\rule{0.400pt}{0.336pt}}
\multiput(635.59,205.98)(0.485,-0.798){11}{\rule{0.117pt}{0.729pt}}
\multiput(634.17,207.49)(7.000,-9.488){2}{\rule{0.400pt}{0.364pt}}
\multiput(642.59,194.50)(0.485,-0.950){11}{\rule{0.117pt}{0.843pt}}
\multiput(641.17,196.25)(7.000,-11.251){2}{\rule{0.400pt}{0.421pt}}
\multiput(649.59,181.89)(0.488,-0.824){13}{\rule{0.117pt}{0.750pt}}
\multiput(648.17,183.44)(8.000,-11.443){2}{\rule{0.400pt}{0.375pt}}
\multiput(657.59,168.03)(0.485,-1.103){11}{\rule{0.117pt}{0.957pt}}
\multiput(656.17,170.01)(7.000,-13.013){2}{\rule{0.400pt}{0.479pt}}
\multiput(664.59,152.55)(0.485,-1.255){11}{\rule{0.117pt}{1.071pt}}
\multiput(663.17,154.78)(7.000,-14.776){2}{\rule{0.400pt}{0.536pt}}
\multiput(671.59,135.55)(0.485,-1.255){11}{\rule{0.117pt}{1.071pt}}
\multiput(670.17,137.78)(7.000,-14.776){2}{\rule{0.400pt}{0.536pt}}
\multiput(678.59,118.64)(0.488,-1.220){13}{\rule{0.117pt}{1.050pt}}
\multiput(677.17,120.82)(8.000,-16.821){2}{\rule{0.400pt}{0.525pt}}
\multiput(686.59,98.60)(0.485,-1.560){11}{\rule{0.117pt}{1.300pt}}
\multiput(685.17,101.30)(7.000,-18.302){2}{\rule{0.400pt}{0.650pt}}
\multiput(693.59,77.37)(0.485,-1.637){11}{\rule{0.117pt}{1.357pt}}
\multiput(692.17,80.18)(7.000,-19.183){2}{\rule{0.400pt}{0.679pt}}
\multiput(700.59,54.77)(0.482,-1.847){9}{\rule{0.116pt}{1.500pt}}
\multiput(699.17,57.89)(6.000,-17.887){2}{\rule{0.400pt}{0.750pt}}
\put(498.0,292.0){\rule[-0.200pt]{1.686pt}{0.400pt}}
\end{picture}

%% file: distributions.tex
\section{Regularization and renormalization as extensions of distributions}
\def\singsupp{\operatorname{singsupp}}

\begin{figure}[htbp]                                 
\begin{center}                                      
\includegraphics[width=0.3\linewidth]{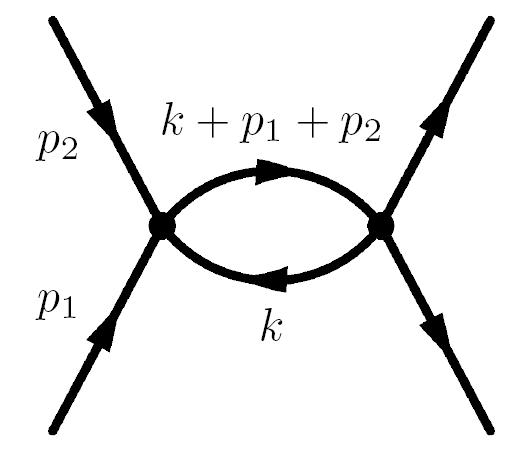}        
\caption[]{Divergent 1-loop diagramm}
\label{feynman1}
\end{center}
\end{figure}

In the previous section, we learned how to make sense of (some)
divergent series of the form $\sum^{\infty}_{k=0} a_{k}=\infty$, but
in QFT the factors $a_k$ are typically complicated mathematical expressions
described by Feynman diagramms, and generically, these expressions
diverge themselves, creating a need for renormalization techniques.
\\
A typical example of a divergent diagramm (in 4 dimensions) is shown
in Figure \ref{feynman1}.

The term described by this diagramm reads (without unimportant factors):
\\
\begin{align}
\int{d^4 k\frac{1}{k^2+m^2}\frac{1}{(p_1+p_2+k)^2+m^2}}\stackrel{k\gg p,m}{\longrightarrow}\int\frac{k^3 dk}{k^4}=\infty
\nonumber
\end{align}
\\
where $m$ denotes the mass of the scalar particles we are scattering. This integral obviously diverges logarithmically for $k\rightarrow\infty$ as shown above. The most straightforward approach to this problem is to introduce a cut-off energy-scale $\Lambda$, such that the divergence at the upper boundary becomes
\begin{align}
\int^{\Lambda}{\frac{k^3 dk}{k^4}}\sim\log(\Lambda),
\nonumber
\end{align}
Usually, such blunt cut-off regularization is incompatible with the
symmetries of the theory at hand and is thus only useful to estimate
``how divergent'' a diagram is (a notion we will below formalize as
the ``singular degree'') and has to be replaced by more sophisticated
methods like dimensional or Pauli-Villars regularization in more
practical applications.

In these notes, instead of momentum representation, we will work in
position space where instead of loop momenta one integrates over the
position of the interaction vertices.

What was $1/(k^2+m^2)$, is now the propagator $G$ defined by the equation
\begin{align}
(\Box+m^2)G(x)=\delta(x),
\label{propagator}
\end{align}

\begin{figure}[htbp]                                 
\begin{center}                                      
\includegraphics[width=0.4\linewidth]{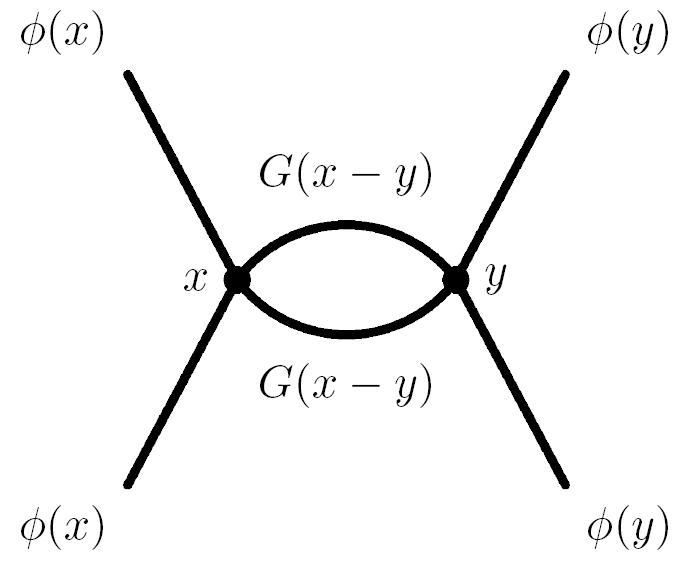}        
\caption[]{Divergent 1-loop diagramm in position-space language}
\label{feynman2}
\end{center}
\end{figure}


we can compute the same diagramm in position space language, which then reads (see Figure \ref{feynman2}): 
\begin{align}
\int d^4 x\int d^4 y\;\phi^{2}_{0}(x)G^2(x-y)\phi^{2}_{0}(y)=\int d^4 x\int d^4 u\;\phi^{2}_{0}(x)G^2(u)\phi^{2}_{0}(x-u)
\label{Gsquare}
\end{align}
The approach of ``causal perturbation theory'' or ``Epstein-Glaser
regularization'' is to take seriously the fact that the propagator is
really a distribution and in the above expression, we are trying to
multiply distributions which in general is undefined. This approach is
advocated in the book by Scharf \cite{Scharf}. Here, we will follow (a
simplified, flat space version) of \cite{Brunetti:2009pn} and in
particular \cite{Prange:1997iy}.

Specifically, in the defining equation (\ref{propagator}) $\delta$ is not
a \textit{function} but a \textit{distribution} (physicists writing
$\delta(x)$ are trying to imply that this is the kernel of the
distribution $\delta$, i.e. that $\delta$ arises by muliplying the
testfunction by a function $\delta(x)$ and then integrating over $x$,
which of course does not exist). Thus, we should interpret $G(x)$ as a
distribution as well (a priori it is only a weak solution of the
differential equation (\ref{propagator})). But as it is in general not
possible to multiply distributions, as we will see later, we do not
have a naive way to obtain ``$G^2$'' as a distribution. In this
chapter, our goal will be to understand renormalization techniques in
terms of distributions. Our route will be led by the question how to
define the product of two distributions that are almost everywhere
functions (which can be multiplied). We will first, therefore,
recapitulate what distributions actually are. Then, we will see in
which cases it is possible to multiply distributions and in which it
is not. This will lead us to the renormalization techniques we are
searching for.


\subsection{Recapitulation of distributions}
Distributions are generalized functions.  Like in many other cases of generalizations this is done via dualization: Starting from an ordinatry function $f$ (in our case locally integrable, that is $f\colon\MR^n\to \MC$ with $\int_K |f|<\infty$ for each compact $K\subset\MR^n$, so that ``divergence of the integral $\int |f|$ at infinity'' is tolerated) one can view it as a linear functional $T_f$ (called a ``regular distribution'') on the functions of compact support via
\begin{equation}
  T_f\colon\phi\mapsto \int f\phi.
\end{equation}
As the map $f\mapsto T_f$ is injective we can use the $T_f$'s to
distinguish the different $f$'s and view $T_f$ in place of $f$. This
suggests to generalize the construction to all linear functionals
$T:\phi\mapsto T(\phi)$ called distributions of which the regular ones
arising from functions $f$ as above are a subset.

Specifically, distributions are defined to be linear and continuous
functionals on the space of \textit{test functions}
$D(\mathbb{R}^{n})=\mathcal{C}^{\infty}_{0}(\mathbb{R}^{n})$ (the
subscript $0$ meaning compact support) equipped with an appropriate
topology that will not concern us here. So formally, we can denote the
distributions to be elements of a space:
\begin{align}
D'(\mathbb{R}^{n})=\left\{T:D(\mathbb{R}^{n})\rightarrow\mathbb{C}\mid\text{$T$ is linear and continuous}\right\}
\nonumber
\end{align}
Besides the regular distibutions $T_f$ encountered above (of which the
function $f$ is called the \textit{kernel}) the typical example of a
\textit{singular distribution} is the $\delta$-distribution: if we
take a given test function $\phi(x)\in D$, then the
$\delta$-distribution is defined to be the functional
$\delta[\phi]=\phi(0)$. This distribution is not regular even though
physicists pretend it to be with a kernel $\delta(x)$ that is so
singular at $x=0$ that $\int \delta(x)=1$ even though it vanishes for
all $x\ne 0$.


Later, we will make use of the fact that distributions can be
differentiated. Using integration by parts in the integral
representation of a regular distribution, we easily obtain
$T_{f'}[\phi]=-T_{f}[\phi']$ which enables us to define the derivative
of a distribution to be $T'[\phi]\equiv-T[\phi']$. Thus we can take
the derivative of a regular distribution $T_f$ even if the kernel $f$
is not differentiable.


The only operation defined on functions that does not directly carry
over to distributions is (pointwise) multiplication
$(f\cdot g)(x)=f(x)g(x)$. Already $L^1_{loc}$ is not closed under
multiplication (recall that in order for a function to be in
$L^1_{loc}$ it must not have singularities that go like $1/x^\alpha$
with $\alpha\ge 1$, a property not stable under multiplication) and in
general the product of distributions is not defined. Of course, as
long as with $f$ and $g$ also $f\cdot g\in L^1_{loc}$ we still have
the regular distribution $T_{f\cdot g}$ and, from a technical
perspective, in this sections, we will deal with the problem to extend
a distribution that can be written as $T_{f\cdot g}$ for a subset of
test functions (those that vanish where $f\cdot g$ is too singular to
be in $L^1_{loc}$) to all test functions.

To this end, for any distribution $T\in D'$ we define the
\textit{singular support} of T ($\singsupp(T)$) as the smallest
closed set in $\mathbb{R}^{n}$ such that there exists a function $f\in
L^{1}_{\text{loc}}$ with $T[\phi]=T_{f}[\phi]$ for all $\phi\in D$
with supp$\phi\cap\singsupp(T)=\emptyset$. For example
$\singsupp(\delta)=\{0\}$, and the corresponding function $f\in
L^{1}_{\text{loc}}$ is simply $f(x)=0$. So the idea behind this
definition is that every distribution can be written as regular
distributions as long as it is only applied to test functions which vanish in
a neighbourhood of 
the distribution's singular support, which enables us to multiply
distributions if we manage to take care of the singular support.


\subsection{Definition of $G^2$ in $D'(\mathbb{R}^{4}\text{\textbackslash}\{0\})$}

Coming back to our concrete field-theoretic problem for a moment, we
now have to answer one important question: what is the singular
support of $G$? Here, we can utilize the fact that $\Box +m^{2}$ is
an \textit{elliptic} operator\footnote{Explaining it without going
  into details, a differential operator which is defined as polynomial
  of $\vec{\partial}$ (with possible coordinate dependent
  coefficients) is elliptic if it is non-zero if we replace
  $\vec{\partial}$ with any non-zero vector $\vec{y}$. In our
  euclidian examples,
  $\Box=\sum^{4}_{i=1}\partial^2_i\rightarrow|\vec{y}|^2>0$ foy any
  non-zero $\vec{y}$} and it can be shown that if two distributions
$T$ and $S$ are related by $\sigma T = S$ with an elliptic operator
$\sigma$, then $\singsupp(T)\subset\singsupp(S)$, so we immediately see
$\singsupp(G)\subset\{0\}$.

It is clear that the singularity of $G(x)$ at $x=0$ corresponds to the
divergence of the momentum-integral at high energies
$\Lambda\rightarrow\infty$, because in order to probe small distances,
short wavelengths which correspond to high momenta are needed,
therefore we speak of UV-divergencies. In this regime, we can set
$m^2\approx0$, and so (\ref{propagator}) simplifies to $\Box
G(x)=\delta(x)\Rightarrow G(x)\sim\frac{1}{x^2}$ for small
$x$.\footnote{The relation
  $\delta(\vec{x})=-\frac{1}{4\pi}\Box\frac{1}{\left|\vec{x}\right|}$
  is well known to hold in 3 dimensions. In general, $\Box
  |x|^{2-n}\propto \delta$ in $n$ dimensions}

Using this, we see that the position-space integral
$\int_{|x|>\frac{1}{\Lambda}} d^4x\, G^2(x)$ again diverges as
log$(\Lambda)$. Because of this divergence $G^2(x)\notin
L^{1}_{\text{loc}}(\mathbb{R}^4)$, so $G^2$ is still not defined as
distribution in $D'(\mathbf{R}^{4})$. Nevertheless, we can use
$G^2(x)$ as kernel of a distribution in
$D'(\mathbf{R}^{4}\text{\textbackslash}\{0\})=\left\{T:D(\mathbb{R}^{4}\text{\textbackslash}\{0\})\rightarrow\mathbb{C}\mid\text{linear
    and continuous}\right\}$ where
$D(\mathbb{R}^{4}\text{\textbackslash}\{0\})$ is the set of test
functions with $\{0\}\notin\text{supp}\phi$.

We now managed to define a distribution $G^2$, but we still have to
extend it from $D(\mathbb{R}^{4}\text{\textbackslash}\{0\})$ to
$D(\mathbb{R}^{4})$. 
Formally, as a linear map, we have to say what values the extension takes
on $D(\MR^4)/D(\MR^4\setminus \{0\})$ which is still an infinite
dimensional vector space. To control this infinity, we will use the
\textit{scaling degree}.


\subsection{The scaling degree and extensions of distributions}

Consider a scaling-map $\Lambda$ acting on test functions: 
\begin{align*}
\mathbb{R}_{>0}\times D(\mathbb{R}^{n})&\rightarrow D(\mathbb{R}^{n})
\\
\left(\lambda\text{,}\phi\right)&\mapsto\phi_{\lambda}(x)\equiv\lambda^{-n}\phi(\lambda^{-1} x)
\end{align*}
The pullback of this map to the space distributions reads
\begin{align*}
\left(\Lambda^{\ast}T\right)[\phi]=T[\phi_{\lambda}]\equiv T_{\lambda}[\phi],
\end{align*}
which for regular distributions gives
\begin{align*}
T_{f,\lambda}[\phi]=\int{\frac{d^n x}{\lambda^n}f(x)\phi\left(\frac{x}{\lambda}\right)}=\int{d^n x f(\lambda x)\phi(x)},
\end{align*}
so power of $\lambda$ in the scaling map acting on test functions is
chosen such that the kernel $f$ transforms in a simple manner without prefactor.
We now define the \textit{scaling degree} ($sd$) of $T\in D'(M\subset\mathbb{R}^n)$:
\begin{align*}
sd(T)=\inf\Big\{\omega\in\mathbb{R}\Big|\lim_{\lambda\searrow0}\lambda^{\omega}
T_{\lambda}=0\Big\}
\end{align*}
To understand this definition, we have to note several properties:
\begin{itemize}
	\item $sd(T)\in[-\infty,\infty[$
	\item For regular distributions $sd(T_{f})\leq0$
	\item $sd(\delta)=n$
	\item $sd(\partial^{\alpha}T)\leq sd(T)+|\alpha|$ with some multi-index $\alpha$
	\item $sd(x^{\alpha}T)\leq sd(T)-|\alpha|$ with some multi-index $\alpha$\footnote{Remember that distributions do not depend on coordinates, only their kernels. Here we used the definition $(x^{\alpha}T)[\phi]\equiv T[x^{\alpha}\phi]$}
	\item $sd(T_1+T_2)=\max\{sd(T_1),sd(T_2)\}$
\end{itemize}
This leads us to the following important theorem:

\begin{theorem}
If $T_0\in D'(\mathbb{R}^n\text{\textbackslash}\{0\})$ is a distribution with $sd(T_0)<n$, then there is a unique distribution $T\in D'(\mathbb{R}^n)$ with $sd(T)=sd(T_0)$ extending $T_0$.
\end{theorem}

The proof of uniqueness is quite easy: We do it by assuming the
existence of two solutions $T$ and $\tilde{T}$ extending $T_0$, and
showing a contradiction. Obviously supp$(T-\tilde{T})=\{0\}$ and from
this it follows that $T-\tilde{T}=P(\partial)\delta$ with some
polynomial $P$. As can be seen from the above notes,
$sd(P(\partial)\delta)\geq n$ and this would be a contradiction to
$sd(T)=sd(\tilde{T})=sd(T_0)<n$. Existence is shown constructively
using a smooth cut-off function $c_\epsilon(x)$ that is 1 outside a
ball of radius $2\epsilon$ and and vanishes in a ball of radius
$\epsilon$. Then we can define 
\begin{equation}
  \label{eq:31}
  T[\phi]=\lim_{\epsilon\searrow 0} T_0[c_\epsilon\phi],
\end{equation}
where one still has to show that the above limit exists in the sense
of distributions.

The theorem above now enables us to uniquely extend distributions of
low scaling degree to the full space $D'(\mathbb{R}^n)$, but what
about distributions with scaling degree $\geq n$? We will solve this
problem in the next section, and afterwards we will be able to return
to our field-theoretic problem of understanding the nature of $G^2$.

But first we have to determine what the scaling degree of the massive propagator $G$, defined by $\delta=(\Box+m^2)G$. We know that $sd(\delta)=n$, and therefore $sd((\Box+m^2)G)=n$ too. If we denote $sd(G)$ by $w$, from the above items it follows that $sd(\Box G)=w+2$, $sd(m^2G)=w$ and therefore $sd((\Box+m^2)G)=w+2$. From this it follows that $w=n-2$ even for the massive propagator.  

\subsection{Case of distributions with high scaling degree}

Considering now a distribution $T_0\in
D'(\mathbb{R}^n\text{\textbackslash}\{0\})$ with $sd(T_0)\geq n$,
uniqueness as in the above theorem does not hold anymore. But if we take
a test function $\phi\in D(\mathbb{R})$ which vanishes of order
$\omega\equiv sd(T_0)-n$ (``\textit{singular order}'') at $x=0$,
i.e. which can be written as
$\phi(x)=\sum_{|\alpha|=\left\lfloor\omega\right\rfloor+1}x^{\alpha}\phi_{\alpha}(x)$
where $\phi_\alpha(0)$ is finite and $\lfloor\omega\rfloor$ denotes the largest integer not bigger
than $\omega$, we can define
$T[\phi]\equiv\sum_{|\alpha|=\left\lfloor\omega\right\rfloor+1}(x^{\alpha}T_0)[\phi_{\alpha}]$. Then
the distribution $x^\alpha T_0$ has scaling degree less than $n$ and
can thus be uniquely extended. 

A general test function can of course be written as a sum of a
function vanishing of order $\omega$ and a polynomial of degree at
most $\omega$ by subtracting and adding the order $\omega$ Taylor polynomial at $x=0$:
\begin{align}
\phi_s(x)\equiv\phi(x)-\sum_{|\alpha|\leq\omega}\frac{x^{\alpha}}{|\alpha|!}\partial^{\alpha}\phi(0)
\label{phis}
\end{align}
This procedure of subtracting the terms leading to divergencies
is the \textit{regularization} in this framework. Since the extended
distribution $T$ beeing applied to $\phi_s$is unique, by linearity, we
still have to define $T$ only on the monomials in $x$ of maximal
degree $\omega$. There is no further restriction on doing this and
this ambiguity in the extension $T$ is what one would have expected:
Changing the value of $T$ on a monomial $x^\alpha$ correspond to
adding a multiple of $\partial^\alpha\delta$ to $T$.

Note well that the arbitratry values of $T[x^\alpha]$ are exactly
those where $T_0[x^\alpha]$ was undefined (divergent in physicists'
parlance) and selecting a certain value corresponds to picking a {\it
  counter term}. procedure known as {\it renormalization}, as formally
infinite values are replaced by finte ones (that have to be fixed by
further physical input like the measurement of the ``physical mass''
or the physical ``coupling constant'').
In the following small sections, we will try out this method in a few
easy, concrete examples. 

\subsubsection{Example in $n=1$}

In order to let our steps so far become clearer, we are going to apply
them to a simple example in $n=1$. In fact, this example shows already
the full regularization and renormalization procedure.

As can be easily checked, the
function $f(x)=\frac{1}{|x|}$ is not an element of
$L^{1}_{loc}(\mathbb{R})$ because of its pole at $x=0$ is not
integrable (it is of course log-divergent), so we can not a priori use it as kernel of a distribution $T_{f}\in D'({\mathbb{R}})$ as we have seen in section 3.1.
But $f(x)\in L^{1}_{loc}(\MR\text{\textbackslash}\{0\})$ and
$sd(T_f)=1=n$, therefore $\omega=0$. This means that for a test
function $\phi(x)$ with $\phi(0)=0$ we can define
$T_f[\phi]\equiv\int{dx\frac{\phi(x)}{|x|}}$ which gives a finite
result: Using l'H\^opital's rule, we see $\lim_{x\to \pm 0}\frac{\phi(x)}{|x|}
= \lim _{x\to\pm 0} \frac{\phi'(x)}{\text{sign}(x)}=\text{finite}$ and thus the integrand is
finite everywhere. This is similar to what we have done in sections 3.2 and 3.3. 
For other test functions, we can again (as in this section above)
define $\phi_s(x)\equiv\phi(x)-\phi(0)$. Afterwards, we write the
general extension for a distribution acting on $\phi$ as
$T_f[\phi]\equiv T_f[\phi_s]+c\phi(0)$ with one arbitrary constant $c$
of our choice.

The careful reader will have realised that there is still a problem
as $\phi_s$ fails to have compact support when $\phi(0)\ne0$ and thus
the integration now diverges at the boundary $x\to\pm\infty$. We will
deal with this problem below but the important observation is that the
divergence in the ultraviolet, that is at small $x$ is cured. 

\subsubsection{Example in $n=4$}

In our field theoretic problem (\ref{Gsquare}) from above, we have
$G^2\sim\frac{1}{x^4}$ in $\mathbb{R}^4$ which is quite similar to the
previous example, as it is the kernel of a distribution
$T_{G^2}=T_{\frac 1{x^4}}\in D'(\MR^4\setminus \{0\})$. Again, we are looking for an
extension. Once more, we have $sd(G^2)=4=n$. Regularization and
renormalization are as in the example above and yield 
\begin{align}
T_{\frac{1}{x^4}}^r[\phi]=\int{d^4 x\frac{\phi(x)-\phi(0)}{x^4}}+c\delta[\phi]
\label{Trenorm}
\end{align}
with $T_{\frac{1}{x^4}}^r\in D'(\mathbb{R}^4)$ and arbitrary $c$. Again, we successfully got rid of the problems at $x=0$ (at the cost of introducing one constant $c$).

\begin{figure}[htbp]                                 
\begin{center}                                      
\includegraphics[width=0.3\linewidth]{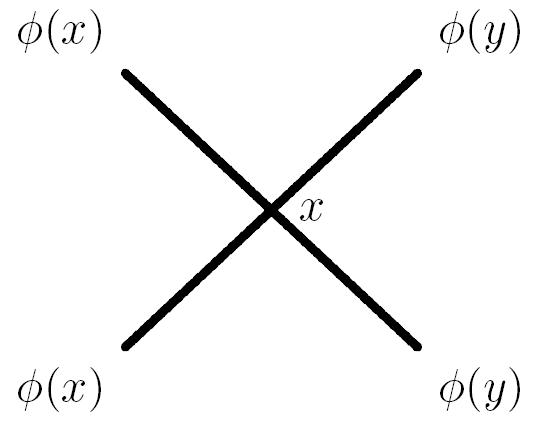}        
\caption[]{Counter-term diagramm.}
\label{feynman3}
\end{center}
\end{figure}

This concludes our calculation of the fish diagram Fig.~\ref{feynman1}
that computes a contribution of the form $\phi(x)^2\phi(y)^2$ to the
effective action of the theory. Since the ambiguous term we found is
$c\delta(x-y)$, the ambiguity in the effective action is indeed
$\phi(x)^4\delta(x-y)$. We see, that it corresponds to the counter
term Fig.~\ref{feynman3} and renormalizes the coupling constant (the
coefficient of the $\phi^4$-term in the action).
\\


\subsubsection{Example with $sd(T)>n$ and preservation of symmetry}

In the two examples above, we had both times distributions $T$ with
$sd(T)=n$ which led us to the introduction of one arbitrary constant
$c$. This amount of ambiguity increases with $sd(T)$, but not all
possible polynomials $P(\partial)\delta$ allowed by the counting can
arise physically. In particular, we require that our theory is still
Lorentz invariant after renormalization and if it has a local gauge
symmetry before that needs to be maintained as well (otherwise one has
an {\it anomaly} that renders the theory ill-defined at the quantum
level since the number of degrees of freedom changes upon
renormalization). 

Let us consider one example where SO(4) invariance (the euclidian
version of the Lorentz group SO(3,1)) selects a subset of the possible
counter terms.

\begin{figure}[htbp]                                 
\begin{center}                                      
\includegraphics[width=0.5\linewidth]{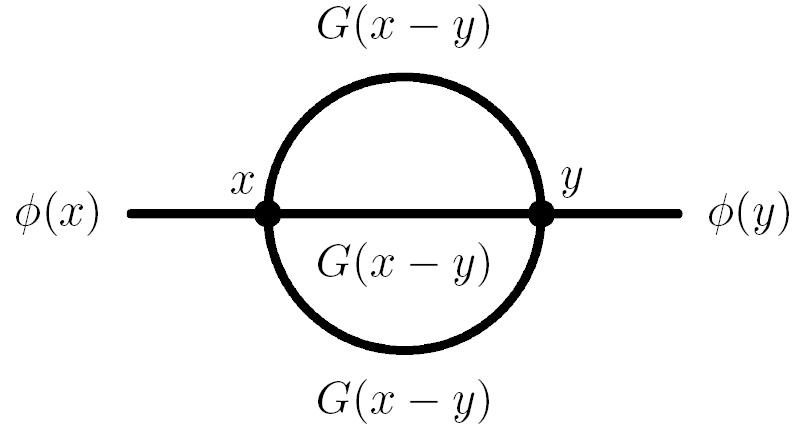}        
\caption[]{Setting sun diagramm in quartic interaction}
\label{feynman4}
\end{center}
\end{figure}

In a theory with potential $\propto \phi^4$ (\textit{quartic
  interaction}), there cannot only be diagramms like Figure
\ref{feynman1}, but also such ones like Figure \ref{feynman4}, known
as the {\it setting sun} diagram.
\\


The term encoded by this diagramm obviously contains
$G^3\sim\frac{1}{x^6}$, which has $sd(G^3)=6>n=4$. By performing the
same steps as above, in this case we a priori get an ambiguity
$c_1\delta+c_2^{i}\partial_i\delta+c^{ij}_3\partial_i\partial_j\delta$
with in total $1+4+\frac{4(4+1)}{2}=15$ arbitrary constants, but upon imposing
SO(4)-invariance this reduces to
$c_1\delta+c_3\Delta\delta$ with only 2 arbitrary constants.

In the effective action, as above, they contribute to the quadratic
terms (as the diagram Fig.~\ref{feynman4} has two external lines)
$\phi(x)(c_1 \delta(x-y)+ c_3
\Delta\delta(x-y))\phi(y)=\phi(x)(c_3\Delta+c_1)\phi(x)\delta(x-y)$. We
recognize that $c_3$ is a {\it wave function renormalization} while
$c_1$ renormalizes the mass-term $m^2\phi^2$.

The fact that $\phi^4$-theory is renormalizable in $n=4$ means that
these two counter terms and the one in the previous subsection are the
only ambiguities that arise when any Feynman diagram of the theory is
renormalized, a proof of being well beyond the scope of these notes.


\subsection{Regaining compact support and RG flow}

In the above calculations, we ignored an important problem: $\phi_s(x)=\phi(x)-\phi(0)$ is not necessarily a test function, as it obviously has $\lim_{x\rightarrow\infty}=-\phi(0)$, therefore for example the integral $\int{dx\frac{\phi(x)-\phi(0)}{|x|}}$ that we encountered in section 3.4.1 may diverge at infinity. We can solve this by introducing a function $w(x)\in D(\mathbb{R}^n)$ with (without loss of generality) $w(0)=1$. We then change the regularized part (i.e. the part without arbitrary constants) of the integral in (\ref{Trenorm}) to 
\begin{align}
T_{\frac{1}{|x|}}[\phi]\equiv\int{dx\frac{\phi(x)-w(x)\frac{\phi(0)}{w(0)}}{|x|}}.
\label{T}
\end{align}
This is a special case of the general formula
\begin{align*}
\phi_s(x)\equiv\phi(x)-\sum_{|\alpha|\leq\omega}\frac{x^{\alpha}w(x)}{|\alpha|!}\left(\partial^{\alpha}\frac{\phi(x)}{w(x)}\right)\bigg\lvert_{x=0},
\end{align*}
which replaces equation (\ref{phis}). Starting from (\ref{T}) we can now write
\begin{align*}
T_{\frac{1}{|x|}}[\phi]=\int{dx\frac{w(x)(\phi(x)-\phi(0))}{|x|}}+\underbrace{\int{dx\frac{(1-w(x))\phi(x)}{|x|}}}_{=T_{\frac{1-w(x)}{|x|}}[\phi]}.
\end{align*}
The second term already is a perfectly fine distribution, the first term can be manipulated in the following way:
\begin{align*}
\int{dx\frac{w(x)(\phi(x)-\phi(0))}{|x|}}=\int{dx\frac{w(x)}{|x|}\int^{x}_{0}du\phi'(u)}
\end{align*}
Now, in the inner integral we can substitute $u=tx$ and afterwards interchange the integrals:
\begin{align*}
\int{dx\frac{w(x)}{|x|}\int^{1}_{0}dt x \phi'(tx)}=\int^{1}_{0}{dt\int dx \;\text{sign}(x) w(x)\phi'(tx)}
\end{align*}
After this, we can substitute $y=tx$ in the inner integral, giving us
\begin{align*}
&\int^{1}_{0}dt\int\frac{dy}{t}w\left(\frac{y}{t}\right)\text{sign}\left(\frac{y}{t}\right)\phi'(y)
\\
&=\int{dy\underbrace{\int^{1}_{0}\frac{dt}{t}w\left(\frac{y}{t}\right)\text{sign}(y)}_{\equiv f(y)}\phi'(y)}
\\
&=T_{f}[\partial\phi]=-(\partial T_{f})[\phi].
\end{align*}
So also the first term is a good distribution. The function $f(y)$
defined above as function of $y$ is well behaved, as $w(x)$ which
enters its definition is a test function, and therefore extremely well
behaved, in particular vanishes for large arguments, so taking $t\to
0$ does not introduce problems.

As an example, let us now set $w(x)=\theta(1-M|x|)$ (or actually a smoothed out version of this non-continuous function)
\begin{align*}
f(y)&=\int^{1}_{0}\frac{dt}{t}\theta\left(1-M\frac{|y|}{t}\right)\text{sign}(y)
\\
&=\int^{1}_{M|y|}\frac{dt}{t}\theta\left(1-M|y|\right)\text{sign}(y)
\\
&=-\ln(M|y|)\theta\left(1-M|y|\right)\text{sign}(y)
\end{align*}
Morally, we regularized our distribution with non-integrable kernel
$\propto\frac{1}{|x|}$ by substituting the derivative of a
distribution with kernel $\propto\log(|y|)$, which is integrable.

In the above calculations, we introduced a mass/energy-scale $M$. It
is now an important question to ask how the distribution changes under
transformations of this scale, i.e. renormalization group (RG)
transformations generated by $M\frac{\partial}{\partial M}$, so called RG flows. We will
now show that it is only the part $const\cdot\delta(x)$, i.e. the part that
is fixed by arbitrary renormalization constants that will change.

First of all, using $\partial_{x}\text{sign}(x)=2\delta(x)$, we see:
\begin{align*}
f'(x)=\frac{-1}{x}\theta\left(1-M|x|\right)\text{sign}(x)-\log(M|x|)\theta\left(1-M|x|\right)2\delta(x)
\end{align*}
Then, we start with:
\begin{align}
M\frac{\partial}{\partial M}T_{\frac{1}{|x|}}[\phi]=M\frac{\partial}{\partial M}\left(\int{dx\frac{(1-w(x))\phi(x)}{|x|}}+T_{-\partial f}[\phi]\right)
\nonumber
\end{align}
Because of $\frac{1-w(x)}{|x|}=\frac{\theta\left(M|x|-1\right)}{|x|}$ in our example, the first term becomes a distribution with kernel
\begin{align}
M\frac{\partial}{\partial M}\frac{\theta\left(M|x|-1\right)}{|x|}=M\delta(M|x|-1).
\label{Mdelta}
\end{align}
The second term in contrast becomes a distribution with kernel
\begin{align*}
M\frac{\partial}{\partial M}(-f'(x))=-M\delta(M|x|-1)+M\frac{\partial}{\partial M}\left[2\log(M|x|)\theta\left(1-M|x|\right)\delta(x)\right].
\end{align*}
The first term of this expression obviously cancels with the contribution from (\ref{Mdelta}), so $M\frac{\partial}{\partial M}T_{\frac{1}{|x|}}$ turns out to be a distribution with kernel:
\begin{align*}
&M\frac{\partial}{\partial M}\left[2\log(M|x|)\theta\left(1-M|x|\right)\delta(x)\right]
\\
&=2\delta(x)\left[\theta\left(1-M|x|\right)-\log(M|x|)M\delta\left(1-M|x|\right)|x|\right]
\\
&=2\delta(x)
\end{align*}
In the last step, we used the presence of the factor $\delta(x)$ (under an integral!) to set $\log(M|x|)|x|=0$ and $\theta\left(1-M|x|\right)=1$. So, under a renormalization group transformation, the distribution changes by $\delta T\propto const\cdot\delta(x)$, that means that a change of energy-scale corresponds to a change of the (at the beginning) arbitrarily selected renormalization coefficients. 
\\

\subsection{What we have achieved in this section}
We have seen a way to recast what looks like divergent Feynman
diagrams as to what looks like distributions for non-integrable
functions. We could then turn these into proper distributions by first
restricting the space of test-functions and then extend them to a full
distribution, possibly at the price of a finite number of undetermined
numerical constants. Those have to be determined by a finite number of
measurements.

In order for the number of introduced parameters for all Feynman
diagrams of the theory to be finite, the scaling degrees of all
appearing distributions in all diagramms have to be below some maximum,
otherwise the theory is not renormalizable.


%% file: conclusions.tex
\section{Summary}
The material in these notes will not be useful for any concrete
calculation in quantum field theory that a physicist might be
interested in. But they might give him or her some confidence that the
calculation envisaged has a chance to be meaningful. 

We tried to present material that is in no sense original but still is
probably not covered in most introductions to quantum field
theory. Hopefully, it helps to refute some of the prejudices against
(perturbative) quantum field theory that mathematically minded people
may have and helps others to better understand how far the hand waving
arguments that we use in our daily work can carry.

In particular, we put our emphasis on two points: Even if the
perturbative expansion is divergent as a power series it can serve two
purposes: The first terms do provide a numerically good approximation
to the true, non-perturbative result and all terms taken together can
indeed recover the full result but only in terms of Borel resummation
rather than as a power series. Second, unphysical infinite momentum
integrals in the computation of Feynman diagrams can be avoided when
properly expressed in terms of distributions. The renormalization of
coupling constants is then expressed as the problem to extend a
distribution from a subspace to all test functions. The language of
distribution theory allows one to avoid mathematically ill-defined
divergent expressions altogether.